\documentclass[12pt]{elsarticle}
\makeatletter
\def\ps@pprintTitle{%
 \let\@oddhead\@empty
 \let\@evenhead\@empty
 \def\@oddfoot{\centerline{\thepage}}%
 \def\@evenfoot{\thepage}
 \let\@evenfoot\@oddfoot}
\makeatother

\usepackage[authormarkup=none]{changes} 
\definecolor{C0}{HTML}{1F77B4} 
\definecolor{C1}{HTML}{FF7F0E}
\definecolor{C2}{HTML}{008000}
\definechangesauthor[color=C0]{R1}
\definechangesauthor[color=C1]{R2}
\definechangesauthor[color=C2]{RA}

\usepackage{hyperref}

\usepackage[T1]{fontenc}
\usepackage{lmodern}
\usepackage{microtype}
\usepackage[margin=1in]{geometry}
\usepackage[]{algorithm2e}
\usepackage{amsmath,amssymb,amsthm,bm}
\usepackage{mathtools}
\usepackage{graphicx}
 \usepackage{algpseudocode} 
\usepackage{subcaption}
\biboptions{sort&compress}
\usepackage{epstopdf}

\usepackage[utf8x]{inputenc}
\usepackage{xcolor}
\AppendGraphicsExtensions{.tif}

\begin{document}

\title{Graph neural networks for simulating crack coalescence and propagation in brittle materials}
\author[auburn]{Roberto Perera}
\author[auburn]{Davide Guzzetti}
\author[auburn]{Vinamra Agrawal\corref{cor1}}
\cortext[cor1]{Corresponding author: vinagr@auburn.edu}
\address[auburn]{Department of Aerospace Engineering, Auburn University, Auburn, AL, USA}

\begin{abstract}

    High-fidelity fracture mechanics simulations of multiple microcracks interaction via physics-based models quickly become computationally {expensive} as the number of microcracks increases.
    This work develops a Graph Neural Network (GNN) based framework to simulate fracture {and stress} evolution in brittle materials due to multiple microcracks' interaction. 
    {The GNN framework is trained on the dataset generated by XFEM-based fracture simulator.}
    Our framework achieves high prediction accuracy {on the test set} (compared to {an XFEM-based fracture } simulator) by engineering a sequence of GNN-based predictions.
    The first prediction stage determines Mode-I and Mode-II stress intensity factors {(which can be used to compute the stress evolution by LEFM)}, the second prediction {stage } determines which microcracks will propagate, and the final {stage} actually propagates crack-tip positions for the selected microcracks to the next time instant. 
    The trained GNN framework is capable of simulating crack propagation, coalescence and corresponding stress distribution for a wide range of initial microcrack configurations {(from 5 to 19 microcracks)} without any additional modification.
    Lastly, the {framework's simulation time shows speed-ups 6x-25x faster compared to an XFEM-based simulator}. 
    {These characteristics, make our GNN framework an attractive approach for simulating microcrack propagation and stress evolution in brittle materials with multiple initial microcracks.}
\end{abstract}


\maketitle

\section{Introduction} \label{Introduction}
    Modeling the initiation, propagation, and interaction of cracks in engineering materials is critical for evaluating their performance and durability across various applications.
    Since its conception, the field of computational fracture mechanics continues to play a significant role in predicting crack initiation, propagation, coalescence and ultimate material failure.
    High-fidelity modeling techniques such as extended finite element method (XFEM) \cite{Belytschko1999XFEM, Belytschko2000RockXFEM, li2018review}, meshfree methods \cite{garg2018meshfree}, scaled boundary finite element method (SBFEM) \cite{SONG1997329, WOLF2000191, SONG2000211, song2018review}, cohesive zone modeling (CZM) \cite{Park2011CZM,schwalbe2012guidelines,yuan2018critical,MOORE201846}, and phase-field modeling (PFM) \cite{francfort1998revisiting,ambati2015review,ambati2016phase} have provided efficient, scalable and reliable means to predict crack behavior in materials.
    A comprehensive review and comparative study of these methods can be found in \cite{egger2019discrete,sedmak2018computational} and references therein.
    
    Despite their success, high-fidelity simulation techniques may require extensive computational resources depending on the type of material, number of cracks, loading configuration 
    and initial crack orientation.
    For instance, modeling microcrack nucleation, propagation and coalescence at the microstructural scale can be very computationally expensive depending on the number of microcracks and microstructural details.
    Understanding material failure at macroscale from microstructure could require simulating prohibitively 
    large number of microcracks.
    For instance, when transitioning from a 2D problem to a realistic 3D microcrack coalescence problem, a $1 \ m^{3}$ domain  may take several CPU-days to solve \cite{HUNTER201987}.
    A possible solution to circumvent the burden of dimensionality 
    involves reduced-order modeling techniques \cite{lucia2004reduced,oliver2017reduced,HUNTER201987}.
    
    Machine learning (ML) methods offer a promising way to develop such reduced order models.
    In the past few years, ML techniques have increasingly found applications in solid mechanics problems.
    Recent works have used ML methods to predict stress hotspots for different crystal configurations, apply tensor decomposition of complex materials, model the behavior of nonlinear materials, and predict complex stress and strain fields in composites \cite{MANGAL2018122, MANGAL20191, Wang2021Tensor, 8967240, HE2021114034, Yangeabd7416}.     
    ML techniques have also been applied to fracture mechanics problems \cite{PANDOLFI2021114078, LIU2020105, HAGHIGHAT2021114012, ZHANG2020112725, SAHA2021113452, osti_1765066}, including 
    simulating 2D crack propagation problems such as predicting structural response due to crack growth, predicting displacement of notched plates, and simulating crack growth in graphene \cite{FENG2021113885, IM2021114030, Lew2021DeepLM}.       
    In one of the recent studies 
    \cite{MOORE201846, HUNTER201987},
    authors first used the high-fidelity model Hybrid Optimization Software Suite (HOSS) \cite{Knight2020HOSS,Euser_2019} to gather a large dataset of microcrack coalescence tests.
    The dataset was then used to train a graph-theory-inspired artificial neural network for predicting connecting cracks, along with an estimated time to failure of the system.
    Another recent study of ML for fracture mechanics implemented a molecular dynamic (MD) simulation model to gather a training dataset, which was then used along a convolutional long-short term memory (ConvLSTM) network for simulating crack growth \cite{HSU2020197}.
    In their approach, the authors demonstrated that by converting the training input to a binary image depicting crack locations as black pixels and the surrounding space as white pixels, the ConvLSTM network was able to predict the next time-step for the crack path as a binary image.   
    Although these studies apply ML to fracture mechanics, predicting dynamic {crack propagation and stress evolution} simulations of higher-complexity problems { with varying number of initial microcracks} has not yet been studied.  
    
    In this work we apply the Graph Network-based Simulator (GNS) to simulate multi-crack dynamics.
    The GNS framework was introduced in 2020 to simulate particle dynamics of fluids and deformable materials when interacting with a rigid body \cite{sanchezgonzalez2020learning}. 
    The GNS approach 
    integrates graph theory along with three ML multi-layer perceptron (MLP) networks.
    The framework was used to represent up to 20k water, sand, and viscous fluid particles as nodes, and their neighboring interactions of energy, momentum, and/or gravitational force as message-passing edges; conceptually similar to nodes and connecting edges in neural networks.
    The information from the message-passing latent space and previous time-steps was used to predict the accelerations and future position of particles, with similar accuracy to conventional numerical integrators used in high-fidelity models.
    This work was extended to develop a mesh-based graph neural network, called \textit{MeshGraphNet} \cite{pfaff2021learning} to simulate finite element simulations of plate bending, flow over an airfoil, flow over a cylinder, and flag dynamics with high accuracy compared to the conventional models.
    Recenly, the GNS approach was applied for simulating the homogenized responses of polycrystals \cite{VLASSIS2020113299}.
    In their work, the authors first developed a graph network representation for polycrystals using vertices for each crystal within the structure, and edges connecting the crystals which shared a face.  
    They then used the GNS approach to predict energy functionals, and simulate anisotropic responses of polycrystals in phase-field fracture.  
    
\begin{figure}
        \centering 
        \includegraphics[width=\linewidth]{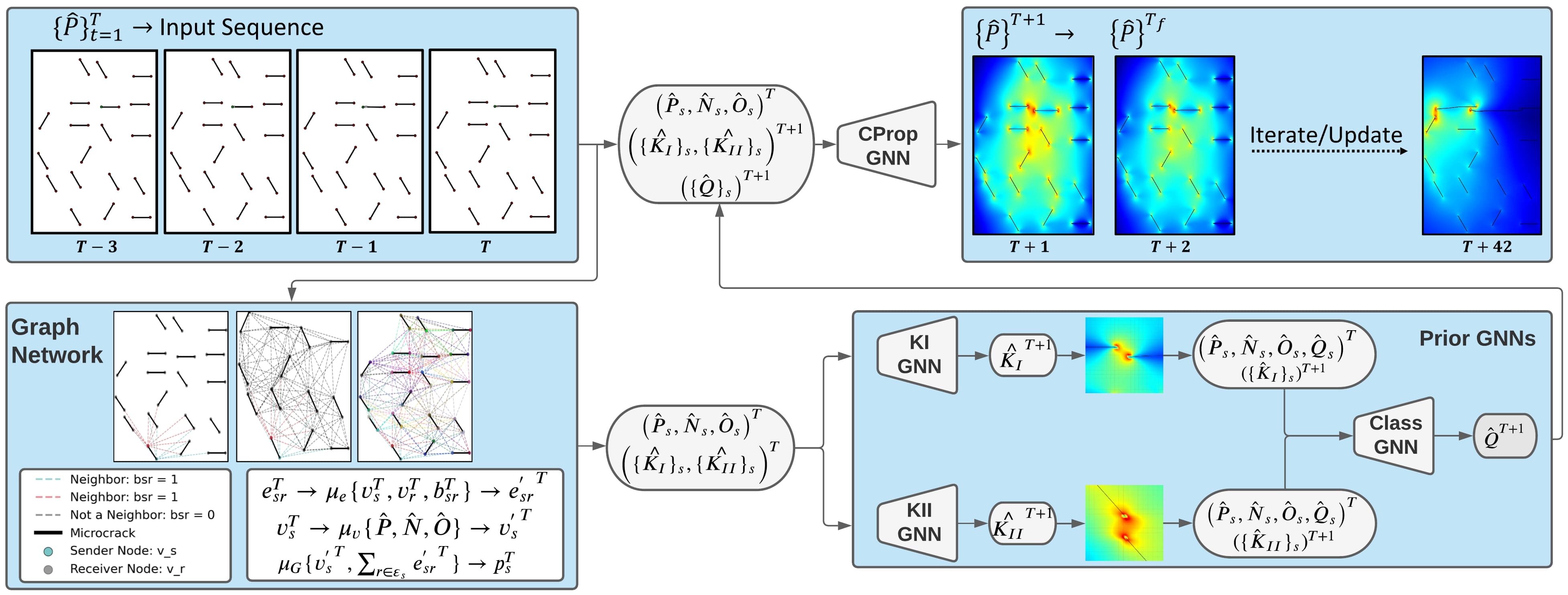}
        \centering
        \caption{Flowchart of the {Microcrack-GNN} framework's structure}
        \label{fig:MicroCrack-GNN_structure}
    \end{figure}

    In our 
    work we leverage advancements in graph neural networks (GNNs) to develop a framework (\textit{Microcrack-GNN}) capable of simulating microcrack propagation and coalescence in brittle materials with high accuracy.
    We considered higher-complexity problems involving multiple microcracks (5 to 19). 
    The structure of the {Microcrack-GNN} framework shown in Figure \ref{fig:MicroCrack-GNN_structure} consists of four GNNs; each GNN is intended to model an underlying physics component of microcrack mechanics.
    In the XFEM approach, propagating crack-tips are determined by their resulting stress distributions \cite{ZHUANG201413}; computed using the superposition of Mode-I and Mode-II effects. 
    We capture this relationship using two GNNs, \textit{$K_{I}$-GNN} and \textit{$K_{II}$-GNN}, predicting Mode-I and Mode-II stress intensity factors respectively. 
    The third GNN, \textit{Class-GNN}, predicts propagating vs non-propagating microcracks, capturing the quasi-static nature of the problem. 
    Using these predictions, the fourth GNN, \textit{CProp-GNN}, predicts the future positions of the crack-tips. 
    By integrating each of these GNNs as shown in Figure \ref{fig:MicroCrack-GNN_structure}, we demonstrate that the framework not only predicts crack coalescence and crack paths with high accuracy, but also stress distributions throughout the domain. 
    We present a new data-driven approach which may serve as a baseline for simulating higher-complexity fracture problems with fewer computational resources and time requirements.
    
    The paper is organized as follows. 
    In Section \ref{sect:Methods}, we describe the XFEM-based model used for gathering training and validation datasets{,} the graph network representation, the generation of {nearest-neighbours}, and the spatial message-passing procedure used for each GNN.
    {In Section \ref{sect:Setup}, we} present a description of the problem set-up including the training-set, validation-set, and test-set, along with the set-up process for varying number of microcracks.
    {In Section \ref{sect:Framework} }, we introduce the structure for the implemented {Microcrack-GNN}.
    This includes detailed descriptions for the {$K_{I}$-GNN}, {$K_{II}$-GNN}, {Class-GNN}, and {CProp-GNN}.
    {In Section \ref{sec:Cross-Validation}, we report the cross-validation results for various training parameters.}  
    {In Section \ref{sec:Results}, we } present the {Microcrack-GNN}'s ability to predict microcrack propagation and coalescence until failure and microcrack length growth for cases involving 5, 8, 10, 12, 15, and 19 initial microcracks, error analyses for the predicted crack length growth, {error analyses for the predicted final crack paths}, {error analyses for the} predicted effective stress intensity factors{, and performance comparison with two additional baseline networks}.  
    Finally, in Section \ref{sec:Results} we compare the required simulation times until failure of the Microcrack-GNN versus the XFEM-based model when increasing the number of initial microcracks (from 5 to 19).

\section{Methods}\label{sect:Methods}

    \subsection{XFEM-based model}\label{subsect:Matlab}
        
        We use the open source XFEM-based model presented in \cite{SUTULA2018205,SUTULA2018225, SUTULA2018257} for simulating various cases of fracture mechanics for brittle materials. 
        This framework (written in MATLAB) is capable of modeling propagation of multiple cracks with arbitrary orientations in a 2D domain.
        Additionally, the XFEM framework is capable of applying various crack growth criteria defined by the user, such as the minimum total energy, the maximum hoop stress, and the symmetric localization criteria. 
        To speed up computation, the authors account only for changes in the fracture topology to compute the timewise force vectors, stiffness matrices, and mesh-enrichment at each crack-tip.
        They then use the domain-form interaction integral approach \cite{GONZALEZALBUIXECH2013129, Zhu2012Improved} to compute the 2D stress intensity factors, $K_{I}$ and $K_{II}$, taking into account the residual strain or stress of each crack, and their surface pressure.
        We use this model to generate a dataset of 2D fracture simulations with multiple microcrack propagation and coalescence.
        
    \subsection{Graph Network Representation}\label{subsect:GraphTheory}
        
\begin{figure} 
            \begin{subfigure}[c]{0.24\textwidth}
                \centering
                \begin{subfigure}[t]{1\textwidth}    
                   \centering \includegraphics[width=0.98\linewidth]{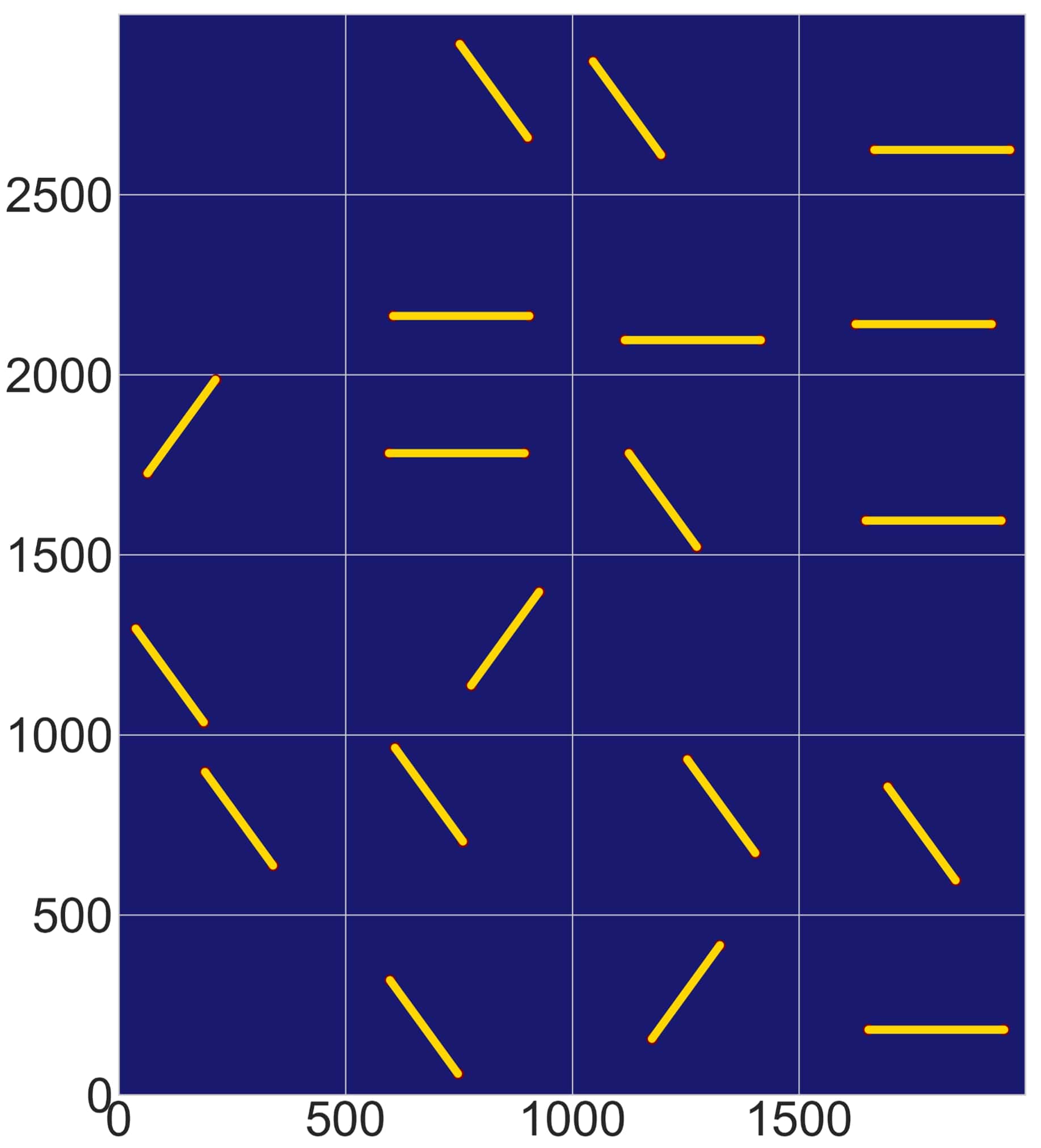}
                    \caption{Initial Configuration}
                    \label{fig:a_neighbors_vertex_edge}
                \end{subfigure}
            \end{subfigure}
            \begin{subfigure}[c]{0.24\textwidth}
                \centering
                \begin{subfigure}[t]{1\textwidth}    
                  \centering
                   \includegraphics[width=0.98\linewidth]{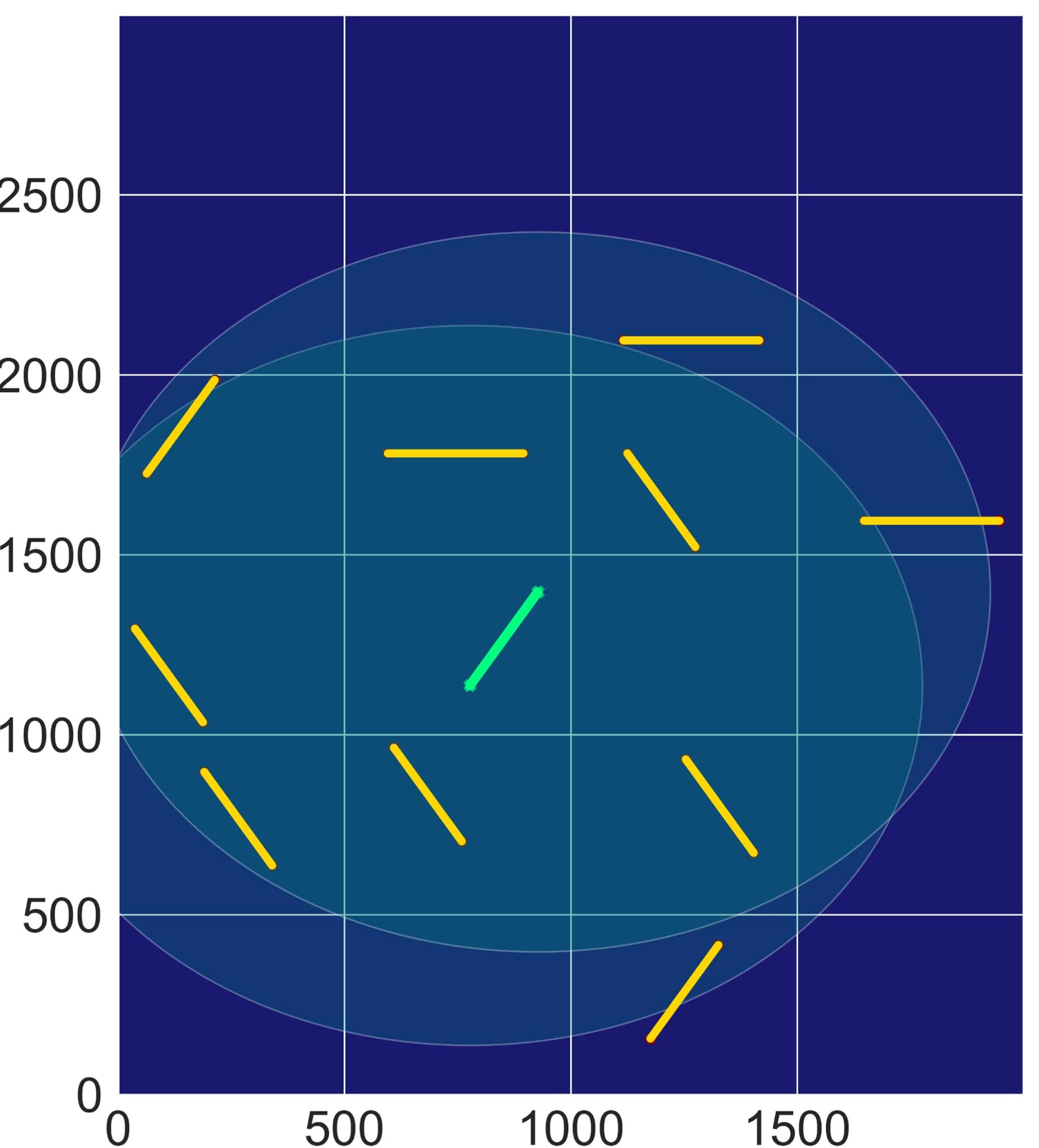}
                    \caption{Crack No.9 Neighbors}
                    \label{fig:b_neighbors_vertex_edge}
                \end{subfigure}
            \end{subfigure}
            \begin{subfigure}[c]{0.24\textwidth}
                \centering
                \begin{subfigure}[t]{1\textwidth}    
                   \centering
                    \includegraphics[width=0.98\linewidth]{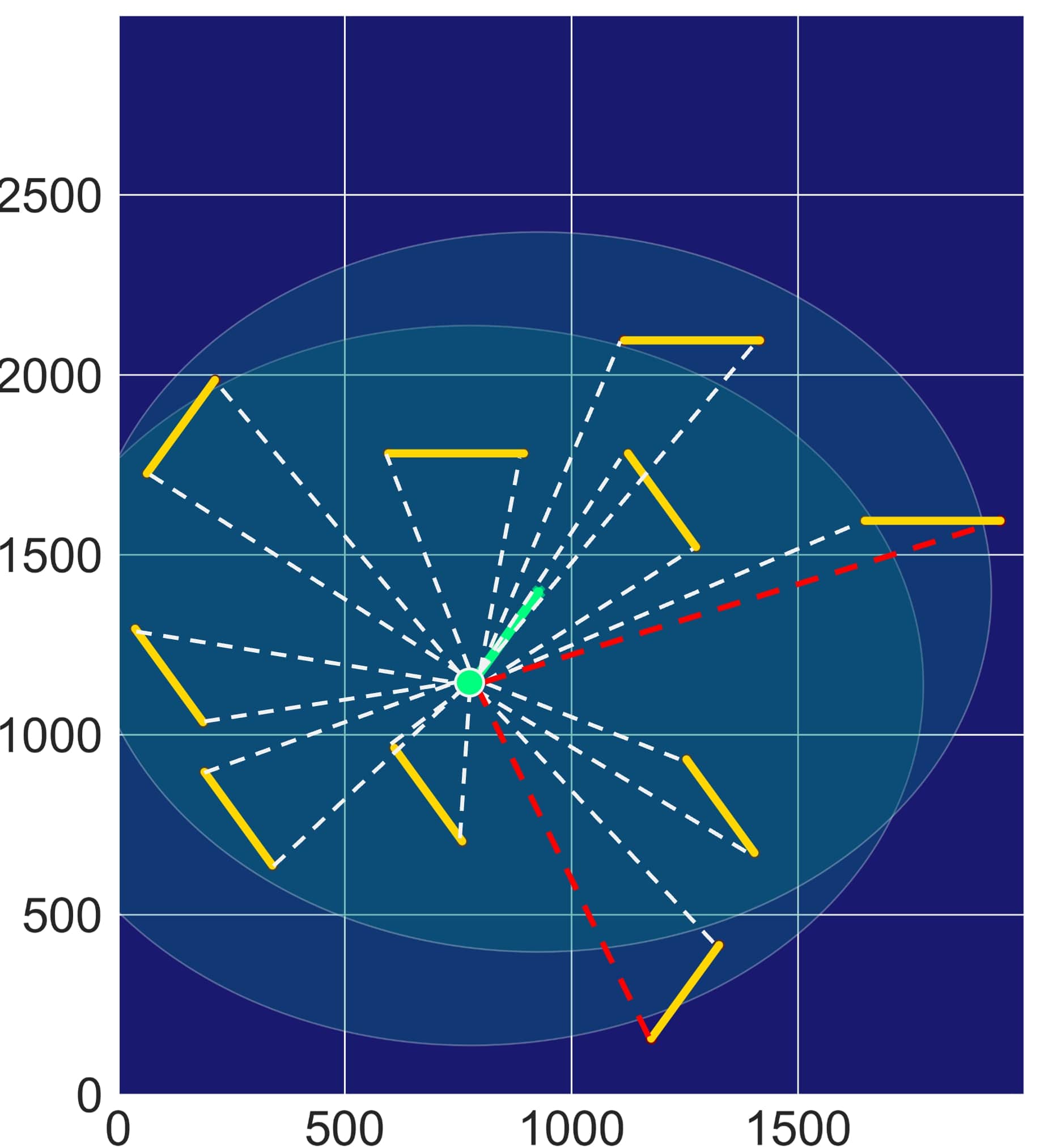}
                    \caption{Vertex No.1 Neighbors}
                    \label{fig:c_neighbors_vertex_edge}
                \end{subfigure}
            \end{subfigure}
            \begin{subfigure}[c]{0.24\textwidth}
                \centering
                \begin{subfigure}[t]{1\textwidth}    
                   \centering
                    \includegraphics[width=0.98\linewidth]{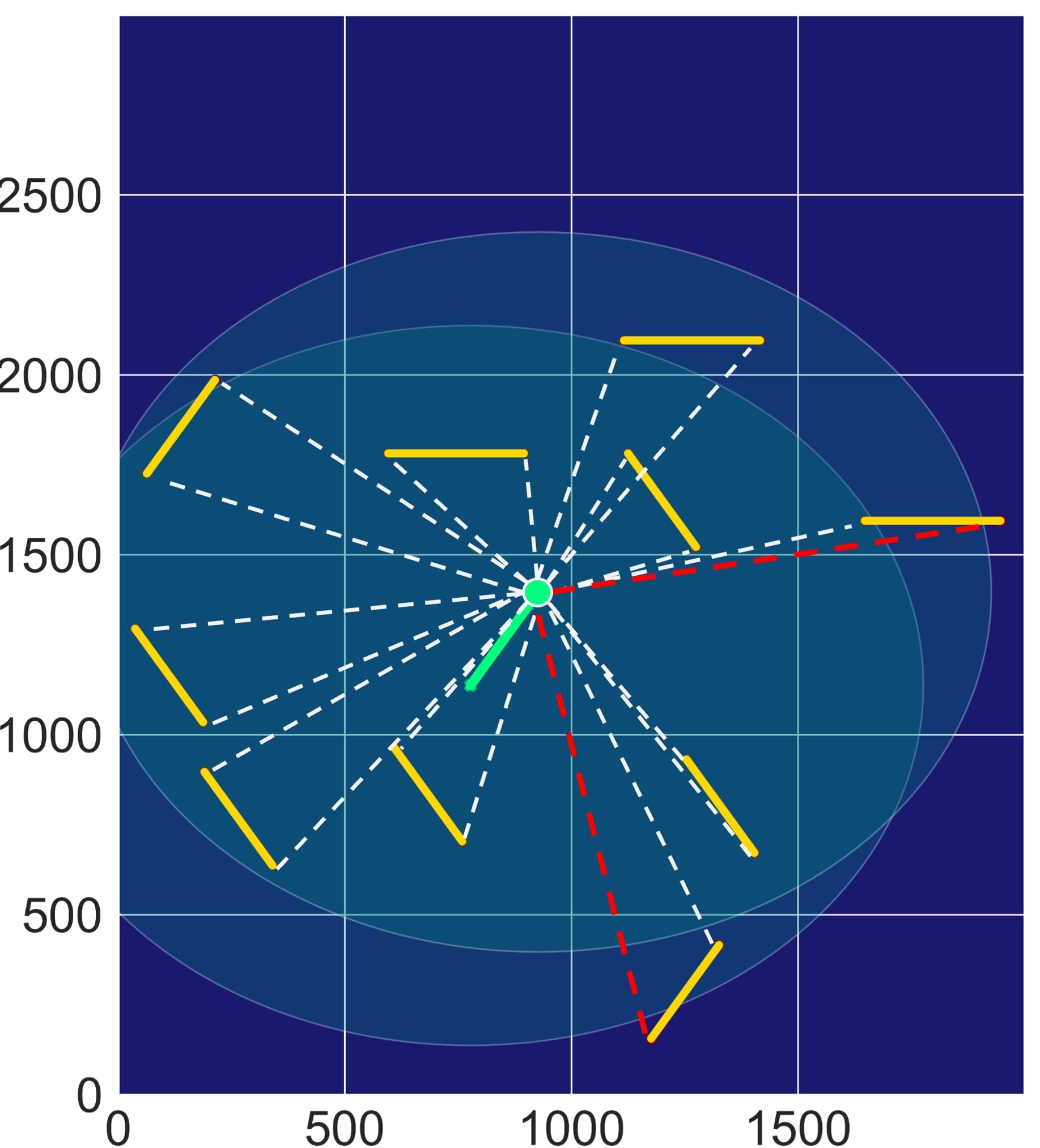}
                    \caption{Vertex No.2 Neighbors}
                    \label{fig:d_neighbors_vertex_edge}
                \end{subfigure}
            \end{subfigure}
            \caption{Representation of current crack-tip (Vertex) connection to neighboring crack-tips (Edges) for a single crack.}
            \label{fig:neighbors_vertex_edge}
        \end{figure}

        In the GNN model, we describe the system as {$\langle \mathbf{V},\mathbf{E} \rangle$}, where $\mathbf{V}$ represents all crack-tips as vertices, and $\mathbf{E}$ represents all the edges in the graph.
        {The edges representation, $\mathbf{E}$, includes edges connecting each crack-tip $v_{s} \in \mathbf{V}$ ({i.e., for any positive integer } $s: \left\{1,2,\ldots, 2C \right\}$ {where $C$ corresponds to the total number of microcracks}) to other crack-tips within a zone of influence (as shown in Figures \ref{fig:a_neighbors_vertex_edge} - \ref{fig:b_neighbors_vertex_edge}), as well as their edges to non-connecting crack-tips (i.e., outside the zone of influence).}
        
        The crack-tip vertices for a sequence of previous time-steps, {$\hat{\mathbf{T}}:= \{T-3, \ T-2, \ T-1, \ T\}$}, are defined by their timewise positions, $\hat{P}_{s}$, their nearest-neighboring crack-tips $\hat{N}_{s}$, as well as their initial orientation $\hat{O}_{s}$ (i.e., {$\{ \theta_{s} \} = $} $0^{o}$, $60^{o}$, {or} $120^{o}$).
        
        {\begin{flalign}
        && \hat{P}_{s} = \{ \left(x_{s}, y_{s} \right)\}_{t=T-3}^{T} && \{s \in \mathbf{V}\}, \nonumber\\ 
        && \hat{N}_{s} = \{ N_{s} \}_{t=T-3}^{T} && \{s \in \mathbf{V}\}, \nonumber\\
        && \hat{O}_{s} = \{ \theta_{s} \}_{t=T-3}^{T} && \{s \in \mathbf{V}\}, \nonumber\\
        && \{v_{s}\}^{\hat{T}} = \left(\hat{P}_{s}, \hat{N}_{s}, \hat{O}_{s} \right) && \{s \in \mathbf{V}\}.  \label{eq:vertex_states}
        \end{flalign}}
        
        Here $(x_{s}, y_{s})$ and $\theta_{s}$ are the Cartesian coordinate positions and orientation (in radians) of vertex $s\in \mathbf{V}$.
        Additionally, at every discretized time-step, the edges in the system will be defined by $(s, r, b_{sr}) \in \mathbf{E}$ arrays, where $s$ is the index for the ``sender'' vertex, $r$ is the index for the ``receiver'' vertex {(i.e., for any positive integer $s: \{1, 2, \dots, 2C \}$}  {and} $r: \{1, 2, \dots, 2C \}${, where $C$ corresponds to the total number of microcracks)}, and $(b \in \{0,1\})$ is a binary value specifying whether the ``sender'' vertex and the ``receiver'' vertex form part of the same pairwise neighbors.
        {We note that when iterating through each crack-tip in the system, we set $b_{sr} = 1$ for $s=r$.}  
        Using this representation, we define a series of neighbors for each microcrack in time sequence {$\hat{\mathbf{T}}$} as
        \begin{flalign}
        && e_{sr}^{t}  = \left(v_{s}^{t}, v_{r}^{t}, b_{sr}^{t}\right) &&  \{t \in \mathbf{\hat{T}}\} \ \ ; \{(s,r,b_{sr}) \in \mathbf{E}\}. \label{eq:discreet_neighbors}
        \end{flalign}
        Here $v_{s}^{t}$ defines the current crack-tip (or ``sender'' {node}) at time $t$, $v_{r}^{t}$ is the neighboring crack-tip (or ``receiver'' {node}) at time $t$, and $b_{sr}^{t}$ is a binary value informing the graph network whether $v_{s}^{t}$ and $v_{r}^{t}$ are within the same neighborhood at time $t$.

    \subsection{Nearest-neighbor sets formulation}\label{subsect:neighbor}   
        Next, we compute all microcrack tips that lie within a zone of influence of {$r_{c} = 750 \ mm$ } as depicted in Figure \ref{fig:c_neighbors_vertex_edge} and \ref{fig:d_neighbors_vertex_edge} by the white-dashed lines.
        Additionally, if a single crack-tip from a neighboring crack, $v_{r}$, falls within the zone of influence of the current crack, (i.e., {$e_{sr} = \left(v_{s}, v_{r}, b_{sr}=1 \right)$}) then the other crack-tip of the neighboring crack is also considered a neighboring vertex.
        In this case, a connecting edge is assigned  {$e_{s(r+1)} = \left(v_{s}, v_{r+1}, b_{s(r+1)}=1 \right)$} as shown in Figure \ref{fig:neighbors_vertex_edge}c by the red-dashed line.
        In a similar fashion, the neighbors of already-coalesced crack-tips are also shared with the neighbors of the connecting crack-tips, and vice-versa.

    \subsection{Spatial Message-Passing Process}\label{subsect:message_passing} 
        The purpose of the spatial message-passing process in GNNs is to learn relationships of the latent space for the vertices, edges, and resulting nearest-neighbors \cite{dwivedi2020benchmark}.
        In this work, the implemented message-passing process involves three main functions to update the system in time.
        
\begin{figure}
            \centering 
            \includegraphics[width=\linewidth]{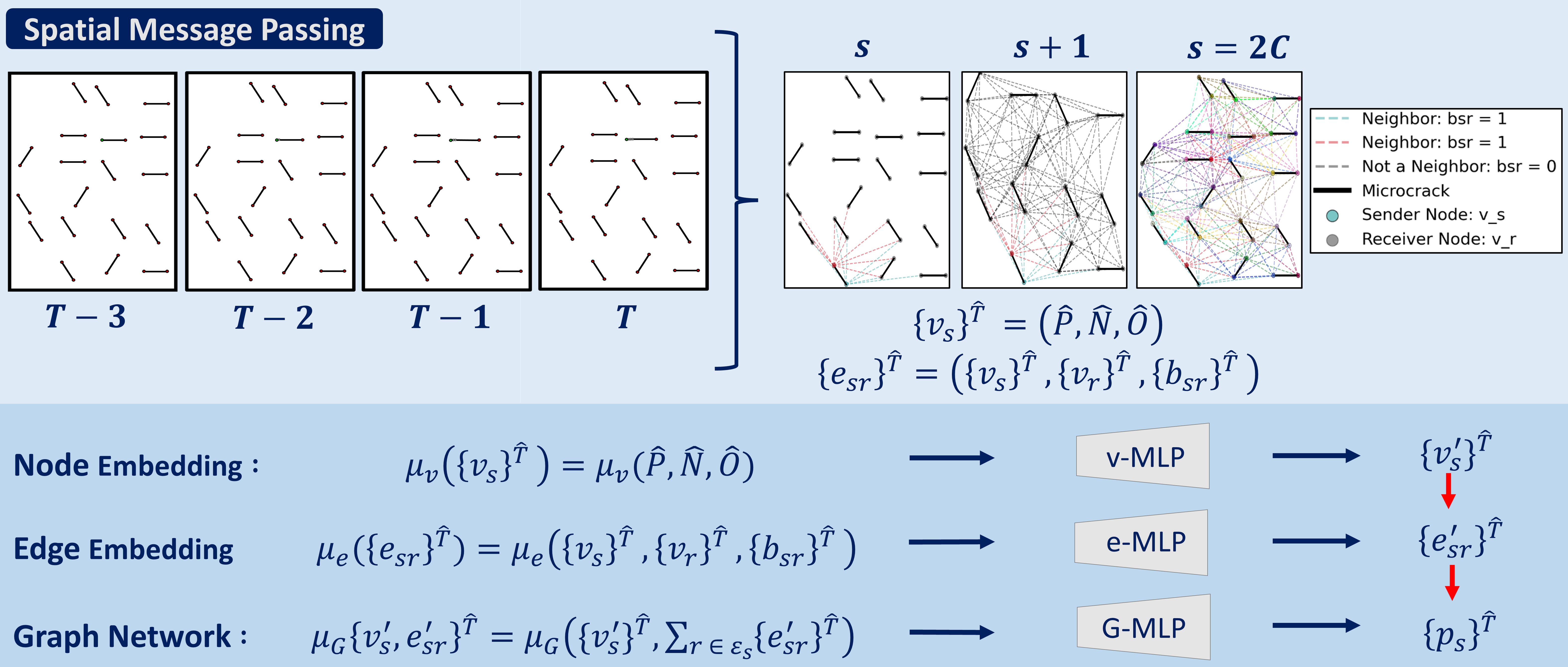}
            \centering
            \caption{Spatial Message-Passing Process. The graph network representation is developed and used with two initial encoder MLPs, "v-MLP" and "e-MLP", to generate the vertices' and edges' embedding in the latent space. For the message-passing stage, the vertices' and edges' latent space embeddings are then concatenated and input to the message-passing MLP network "G-MLP", to propagate the vertex-edge interactions for a series of update steps, M. The final output is one-hot encoded feature vector describing the vertex-edge-vertex system interaction in latent space.}
            \label{fig:Message_Passing}
        \end{figure}

        First, we develop the graph network representation for the vertices (crack-tips) and their feature vectors, as well as their respective nearest-neighbor configuration as described 
        in Section \ref{subsect:GraphTheory} using equation (\ref{eq:vertex_states}). 
        We then pass the resulting feature vector for the vertices as input to an encoder MLP, denoted as ``v-MLP'' in Figure \ref{fig:Message_Passing}, and represented by $\mu_{v}$ in equation (\ref{eq:edges_MLP}a).
        The outputs from the vertices' MLP encoder network, {$\{v_{s}^{'}\}^{\hat{T}}$}, are the encoded vertices' embedding in the latent space for a sequence of time {$\hat{\mathbf{T}}$}.
        
        Second, we learn the relationship effects between vertex-edge-vertex interactions.
        We achieve this by applying an additional MLP encoder network to the resultant edges embedding obtained from equation (\ref{eq:discreet_neighbors}).
        The edges MLP encoder network is denoted in Figure \ref{fig:Message_Passing} with ``e-MLP'', and depicted in equation (\ref{eq:edges_MLP}b) as $\mu_{e}$. 
        As a result, the outputs from equation (\ref{eq:edges_MLP}b), {$\{e_{sr}^{'}\}^{\hat{T}}$}, are the encoded edges' embedding describing the vertex-to-edge interactions in the latent space for the time sequence {$\mathbf{\hat{T}}$}.
        
       \begin{flalign}
        && \{v_{s}^{'}\}^{\hat{T}} \longleftarrow \mu_{v}\left(\hat{P}_{s}, \hat{N}_{s}, \hat{O}_{s} \right) && \{s \in \mathbf{V}\} \  \nonumber\\
        && \{e_{sr}^{'}\}^{\hat{T}}  \longleftarrow \mu_{e}\left(\{v_{s}\}^{\hat{T}}, \{v_{r}\}^{\hat{T}}, \{b_{sr}\}^{\hat{T}}\right) &&  \{(s,r) \in \mathbf{E}\} \  \label{eq:edges_MLP}
        \end{flalign}
        
        Lastly, we concatenate the encoders' output for both the vertices' embedding and for the vertex-edge-vertex embedding and pass it through the message-passing network. 
        The message-passing MLP network, denoted by ``G-MLP'' in Figure \ref{fig:Message_Passing}, allows the network to learn information from the latent space between the crack-tips and edge features, and their interaction \cite{klicpera2020directional, li2017gated, zhang2020dynamic,gilmer2017neural}. 
        To perform the message-passing phase, a series of update steps $M$ are taken on the message-passing MLP.
        We note that depending on the complexity of the problem, the required number of update steps may vary.
        Following a similar approach as presented in \cite{Scarselli2009Graph, battaglia2018relational, battaglia2016interaction}, we {initially} chose four update steps for all GNNs presented in this work.
        {However, after cross validation (Section \ref{sec:Cross-Validation}) the optimal number of message passing steps obtained was $M=6$.}
        As a result a one-hot encoded feature vector describing the latent space vertex-edge-vertex interactions, {$\{p_{s}\}^{\hat{T}}$}, is obtained as  
        \begin{flalign}
        && \{p_{s}\}^{\hat{T}}  \longleftarrow \mu_{G}\left(\{v_{s}^{'}\}^{\hat{T}}, \sum_{r \in \mathcal{E}_{s}} \{e_{rs}^{'}\}^{\hat{T}}\right) &&  \{s \in \mathcal{E}\}. \label{eq:one-hot-encoded}
        \end{flalign}
        Here $\mu_{G}$ is the message-passing MLP network, $\mathcal{E}$ represents all nearest-neighbors of all microcracks, and $\mathcal{E}_{s}$ describes all nearest-neighbors of microcrack $v_{s}$ (i.e., at $b_{sr} = 1$).

\section{Simulations set-up}\label{sect:Setup}   
    
   \subsection{Training-set and Validation-set}\label{subsect:dataset}
        As mentioned in Section \ref{subsect:Matlab}, we used the XFEM-based model to generate the training-set, validation-set, and the test-set.
        The problem set-up was inspired by the work presented in \cite{HUNTER201987}, where a domain of $2000 \ mm$ by $3000 \ mm$ with a maximum of 19 microcracks was used for each simulation. 
        We restricted our analysis to an isotropic, homogeneous and perfectly brittle material.
        We chose the Young's Modulus of $E = 22.6 \ GPa$, with Poisson's ratio of $\nu = 0.242$, and material toughness of $K_{crt} = 1.08 \ MPa \cdot \sqrt{m}$.
        Further, we assumed quasi-static loading and restrict our analysis to neglect crack-tip bifurcation and multiple cracks propagating at the same time.

        Next, we fixed the bottom edge of the domain and apply a constant amplitude of $0.01\ m$ at the top edge {towards the positive y-direction (tensile load perpendicular to top edge)}. 
        We implemented a new function (\textit{$GenCrack\_Rand$}) to generate  user-defined number of cracks $C$ ($5 \leq C \leq 19 $) in random positions and orientations {($0^{o}$, $60^{o}$, and $120^{o}$)} without overlap.
        Using this, we generate a dataset of $64$ simulations for each $C$, resulting in a total of $ N_{sim} = 64\times 15 = 960$ simulations.
        The number time-steps used in the training-set and validation-set was set to $N_{steps} = 46$. 
        Following \cite{sanchezgonzalez2020learning, battaglia2018relational,li2020visual, pfaff2021learning}, we chose the sequence length of $N_{seq} = 4$ for the GNNs while training, that is, {$\mathbf{\hat{T}} := \{T-3, \ T-2, \ T-1, \ T\}$ }.
        In essence, the framework inputs the four initial microcracks' states, and predicts the fifth (future) microcracks' states.
        
        Next, we split the total dataset of $N_{sim} = 960$ simulations as $90\%$ for the training-set ($864$ simulations), and $10\%$ for the validation-set ($96$ simulations).
        We can calculate the total number of inputs for the training and validation set as $N_{input} = N_{sim}\times (N_{steps}-N_{seq})$.
        This resulted in a total of $36,288$ inputs for the training-set, and $4,032$ inputs for the validation set.
        Lastly, we separate the training-set into shuffled batches of size $32$ and kept the validation set in sequential order for batch size of $1$.
        Once the training and validation process was completed, we implemented the same approach for the test set.
        For each $5\leq C\leq 19$, we performed {$15$} XFEM simulations resulting in a total of {$225$} simulations to test the {Microcrack-GNN} framework.
        {Lastly, we emphasize that each simulation can contain between 50-100 time-steps until failure, resulting in a total test set size of up to {22,500} discreet time-steps; predictions are made at each time-iteration.}

    \subsection{Varying number of microcracks}\label{subsect:VaryingCracks}  
        A key feature of the developed framework is its capability to handle different number of microcracks ($C$) between test cases.
        In this work, we allowed the {Microcrack-GNN} framework to handle $5\leq C\leq 19$.
        To overcome the problem in our set-up of varying input-to-output size, we fixed the inputs and outputs of the network at the maximum number of cracks, $C_{max} = 19$.
        We achieved this by including an additional function to count the number of microcracks from the given initial configuration, and assign zeros to the remaining inputs when $C < C_{max}$. 
        This approach showed to maintain efficient training for cases of varying number of microcracks.
        Lastly, we note that this procedure can be easily extended for a higher or lower number of cracks.

\section{Microcrack-GNN Framework}\label{sect:Framework}

    As mentioned in previous sections, the developed framework for predicting multiple microcracks' propagation, interaction, and coalescence evolution involves three initial GNNs ($K_{I}$-GNN, $K_{II}$-GNN, and Class-GNN) as the first step.
    The second step, is to use the generated predictions from the initial GNNs as the input to an additional final GNN (CProp-GNN) for simulating future crack-tips' positions.
    We present the implementation steps for $K_{I}$-GNN, $K_{II}$-GNN, Class-GNN, and CProp-GNN shown in Figure \ref{fig:MicroCrack-GNN_structure} in detail in the following sections.
 
    \subsection{$K_{I}$ - GNN}\label{subsect:KI-GNN}
        
        We implemented the {$K_I$-GNN} to predict the Mode-I stress intensity factors for all crack-tips, $(K_{I})_{s}$, at future time-steps.
        First we generated the input graph representation following the same procedure described in Sections  \ref{subsect:GraphTheory}, \ref{subsect:neighbor}, and  \ref{subsect:message_passing}. 
        The resulting input graph is described as
        \begin{flalign}
            &&(\hat{K}_{I})^{T+1}  \longleftarrow \psi_{K_{I}}(\hat{P}, \hat{N}, \hat{O}, \hat{K}_{I}),\nonumber
            \\
            &&(\hat{K}_{I})_{s}^{T+1}  \longleftarrow \psi_{K_{I}}\left[\{ p_{s} \}_{t=T-3}^{T}|\hat{N}_{s}, \hat{O}_{s}, \{(\hat{K}_{I})_{s}\}_{t=T-4}^{T}\right] &&  \{s \in \mathbf{V}\}.
            \label{eq:discreet_KI}
        \end{flalign}
        Here the first set of inputs are the one-hot encoded vertex-edge-vertex feature vectors from equation (\ref{eq:one-hot-encoded}) along with their groups of nearest-neighbors in time sequence {$\hat{\mathbf{T}}$}.
        The second set of inputs are the initial orientations of all microcracks in the system $\hat{O}$, followed by the Mode-I stress intensity factors $\hat{K_{I}}$ during the four previous time-steps from the time sequence {$\hat{\mathbf{T}}$}.
        Additionally, $\psi_{K_{I}}$ defines the integrated ML MLP network used to train the {Microcrack-GNN} framework for predicting the future Mode-I stress intensity factors at each crack-tip $(\hat{K}_{I})_{s}^{T+1}$.

        Lastly, we used the predicted Mode-I stress intensity factors from the {$K_{I}$-GNN} to calculate the LEFM stress distribution in the domain.
        For this, we first discretized the domain into 201 points in the horizontal and vertical directions resulting in a total of $40,401$ points in the domain. 
        {Then, using the principle of superposition from LEFM \cite{Bower1991Applied}, we computed the Von Mises stresses due to the predicted Mode-I stress intensity factors at each point in the domain.}
        Figure \ref{fig:K1_zoomed} shows the comparison of von Mises stress distribution as predicted by the $K_I$-GNNs against that obtained from XFEM simulations.

    \subsection{$K_{II}$ - GNN} \label{subsect:KII-GNN}
    
\begin{figure}
            \centering 
            \begin{subfigure}[c]{0.48\textwidth}
                \centering
                \includegraphics[width=\linewidth]{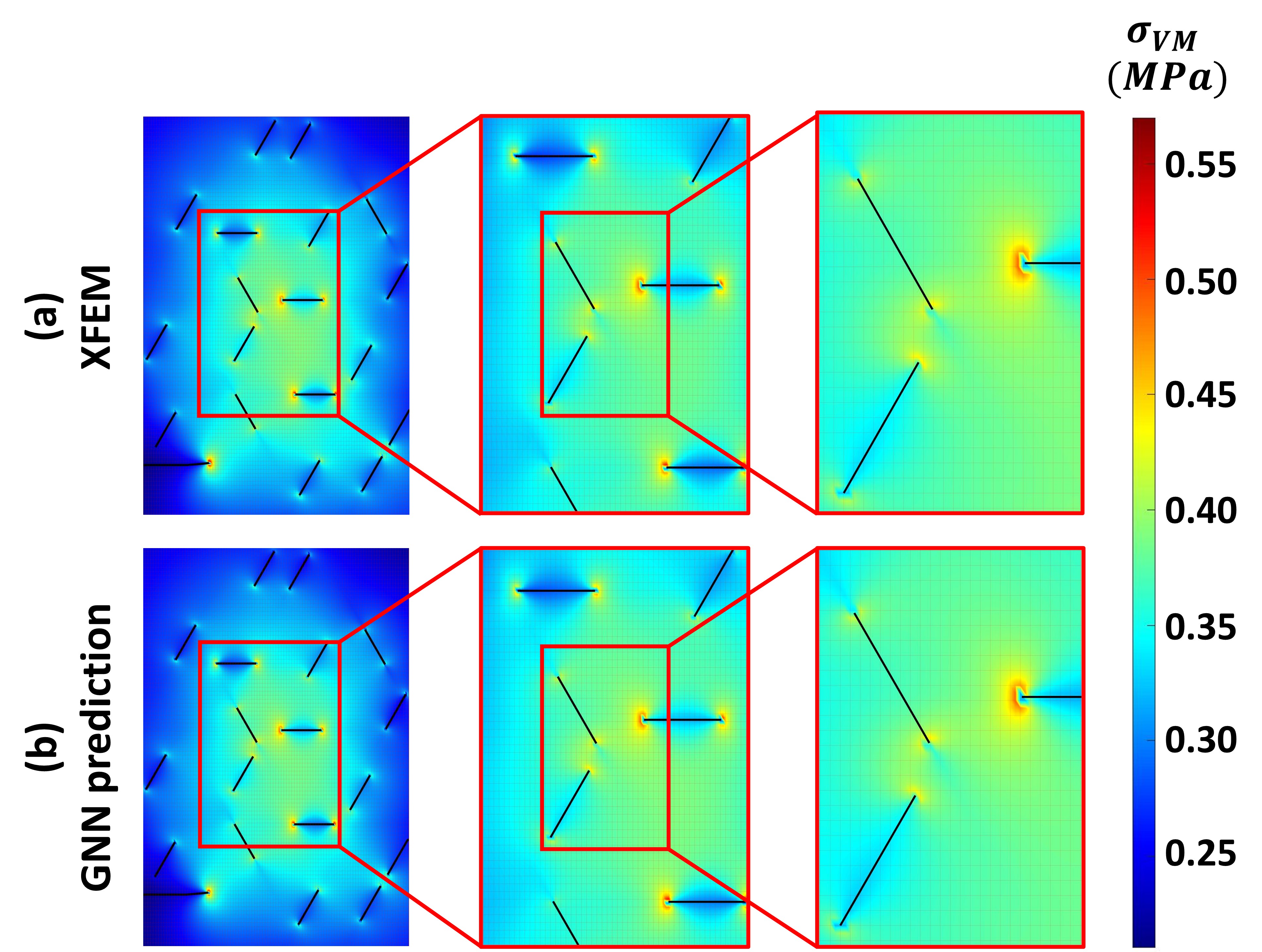}
                \centering
                \caption{$\sigma_{VM}$ from $K_{I}$-GNN predictions}
                \label{fig:K1_zoomed}
            \end{subfigure}
            \centering
            \begin{subfigure}[c]{0.48\textwidth}
                \centering
                \includegraphics[width=0.98\linewidth]{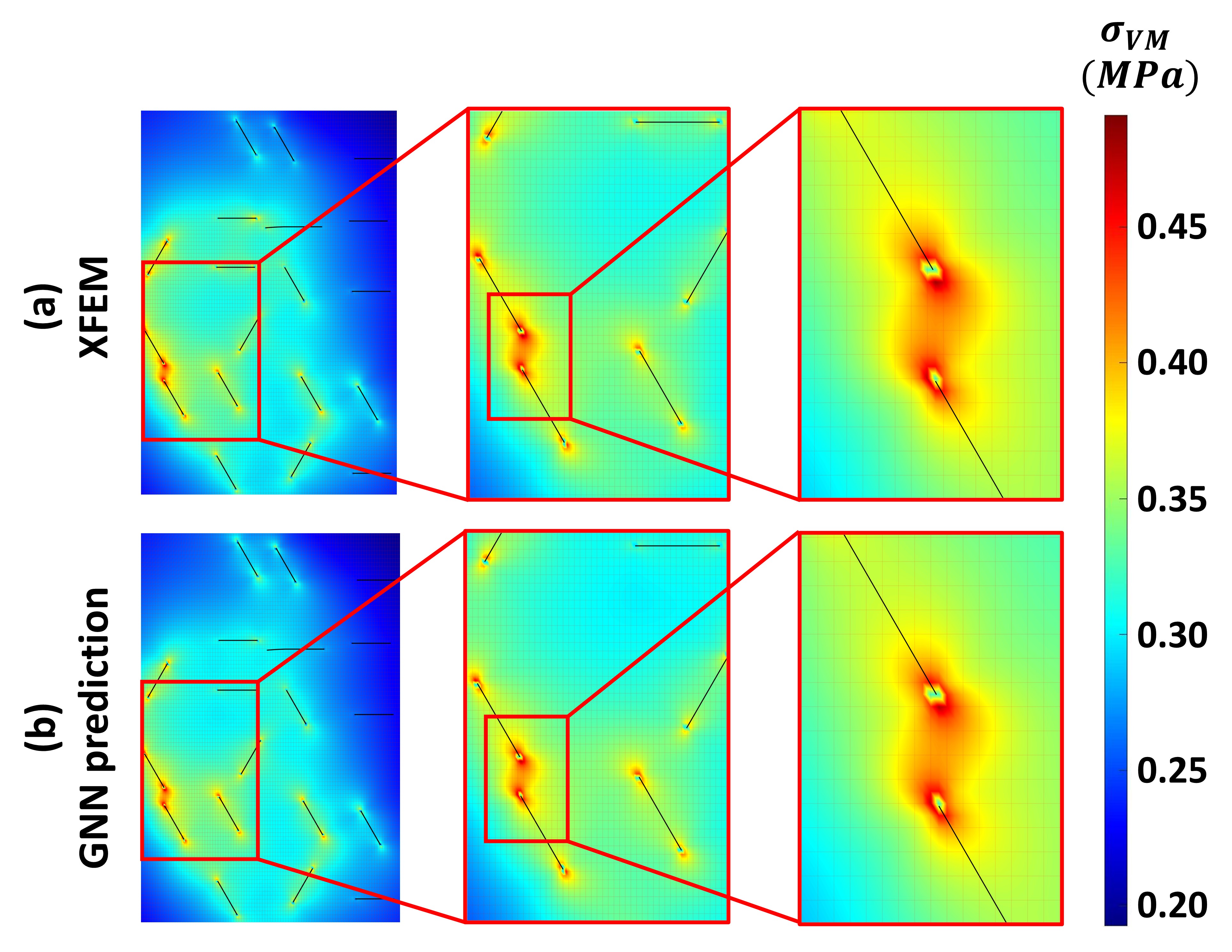}
                \centering
                \caption{$\sigma_{VM}$ from $K_{II}$-GNN predictions}
                \label{fig:K2_zoomed}
            \end{subfigure}
            \caption{Von Mises stress distributions for a) the predicted Mode-I stress intensity factors from the {$K_{I}$-GNN} model and b) predicted Mode-II stress intensity factors from the {$K_{II}$-GNN} model.}
        \end{figure}

        We implemented the {$K_{II}$-GNN}, shown in Figure \ref{fig:MicroCrack-GNN_structure} to predict the Mode-II stress intensity factors for each crack-tip at future times.
        The input graph representation for this model follows a similar structure as for the {$K_{I}$-GNN}, as  
        
        \begin{flalign}
            &&(\hat{K}_{II})^{T+1}  \longleftarrow \psi_{K_{II}}(\hat{P}, \hat{N}, \hat{O}, \hat{K}_{II})\nonumber\\
            && (\hat{K}_{II})_{s}^{T+1}  \longleftarrow \psi_{K_{II}}\left[\{ p_{s} \}_{t=T-3}^{T}|\hat{N}_{s}, \hat{O}_{s}, \{(\hat{K}_{II})_{s}\}_{t=T-4}^{T}\right] &&  \{s \in \mathbf{V}\} \
            \label{eq:discreet_KII}
        \end{flalign}
        Here $\psi_{K_{II}}$ defines the ML MLP network used to train the {Microcrack-GNN} framework for predicting future Mode-II stress intensity factors at each crack-tip $(\hat{K}_{II})_{s}^{T+1}$.
        Lastly, we follow the similar methodology as $K_I-GNN$ and use LEFM equations {\cite{Bower1991Applied}} to compute von Mises stress and compare against XFEM results as shown in Figure \ref{fig:K2_zoomed}. 
        
    \subsection{Classifier - GNN}\label{subsect:Class-GNN}
    
        Next, we concatenated the predicted Mode-I and Mode-II stress intensity factors and used it as input to the {Class-GNN} (Figure \ref{fig:MicroCrack-GNN_structure}).
        The purpose of the {Class-GNN} is to predict a binary feature array, $\hat{Q} \in \{0,1\}$, which defines the propagating crack-tips as $\hat{Q}_{s} = 1$, and non-propagating crack-tips as $\hat{Q}_{s} = 0$.
        The {Class-GNN} captures the quasi-static nature of the problem (one of the assumptions in the XFEM framework) where only one crack-tip is propagating at any given time.
        This also simplified the challenges in training the Microcrack-GNN framework where instead of predicting the future position of all the crack-tips at any given time, the framework only has to predict one crack-tip position at a given time-step.
        
        In the latent space, the {Class-GNN} is intended to learn the relationships that exist for connecting and non-connecting crack-tips (e.g., connecting crack-tips do not propagate further), and propagating and non-propagating crack-tips, prior to the final prediction of  positions.
        The {Class-GNN} uses the binary feature vector of propagating crack-tips from the previous time-step $(\hat{Q})^{T}$, and the predicted $(\hat{K_{I}})^{T+1}$ and $(\hat{K_{II}})^{T+1}$ as the input.
        The CLASS-GNN is expressed as
        \begin{flalign}
            &&(\hat{Q})^{T+1}  \longleftarrow \psi_{CLASS}\left[ (\hat{P}, \hat{N}, \hat{O}) , \hat{Q}^{T}, (\hat{K}_{I}, \hat{K}_{II})^{T+1} \right],   
            \nonumber\\
            && (\hat{Q})_{s}^{T+1}  \longleftarrow \psi_{CLASS}\left[\{ p_{s} \}_{t=T-3}^{T}|\hat{N}_{s}, \hat{O}_{s}, \{\hat{Q}_{s}\}^{T}, \{(\hat{K}_{I})_{s}, (\hat{K}_{II})_{s}\}^{T+1}\right] &&  \{s \in \mathbf{V}\}. \
            \label{eq:discreet_class}
        \end{flalign}
        Here $\psi_{CLASS}$ defines the MLP network used to train the classifier network.
        In other words, the stress intensity factors at future time-steps involve information of system's stress distribution and energy release rate which allow the {Class-GNN} to recognize propagating crack-tips.
        
        We emphasize that we implemented the {Class-GNN} for this specific problem due its quasi-static nature.
        In recent works where GNNs have been developed for prediction of particle dynamic simulations \cite{sanchezgonzalez2020learning, pfaff2021learning, Scarselli2009Graph, klicpera2020directional, battaglia2016interaction}, all particles in the system are changing with respect to time, which is not the case in quasi-static problems.  
        In the case where more than one crack-tip is changing with time (simultaneous crack growth), the {Class-GNN} may not be necessary, or an extension for the network for handling multiple classes (multi-class classification, and/or multi-label classification) \cite{VENKATESAN2016310} may be necessary.

    \subsection{Propagator - GNN} \label{subsect:Cprop-GNN}
    
        The final component of the {Microcrack-GNN} framework is the {CProp-GNN}.
        The purpose of the {CProp-GNN} was to predict the future positions for all crack-tips, given their four previous configurations, and the predictions from the initial GNNs ($K_{I}$-GNN, $K_{II}$-GNN, and Class-GNN).      
        Additionally, as shown in Figure \ref{fig:MicroCrack-GNN_structure} and equation (\ref{eq:discreet_cprop}), the remaining structure of the input graph representation used to train the {CProp-GNN}, consisted of the predicted one-hot encoded vertex-edge-vertex features in the latent space from the message-passing model, and the initial orientations of all microcracks as described in Sections \ref{subsect:GraphTheory} - \ref{subsect:message_passing}.

        \begin{flalign}
            &&(\hat{P})^{T+1}  \longleftarrow \psi_{CProp}\left[ (\hat{P}, \hat{N}, \hat{O}), (\hat{K}_{I}, \hat{K}_{II}, \hat{Q})^{T+1} \right]   
            \nonumber\\
            && (\hat{P})_{s}^{T+1}  \longleftarrow \psi_{CProp}\left[\{ p_{s} \}_{t=T-3}^{T}|\hat{N}_{s}, \hat{O}_{s}, \{(\hat{K}_{I})_{s}, (\hat{K}_{II})_{s}, \hat{Q}_{s},\}^{T+1}\right] &&  \{s \in \mathbf{V}\} \
            \label{eq:discreet_cprop}
        \end{flalign}
        
        By combining the prior GNNs along with the final {CProp-GNN}, the entire flowchart of the {Microcrack-GNN} framework was capable of accurately simulating multiple cracks propagation and coalescence using information from the past configurations.
        Using Microcrack-GNN, we simulated the crack growth process for varying number of cracks.

\section{Cross-Validation}\label{sec:Cross-Validation}
{

    For additional tuning of the GNN, we performed cross-validation \cite{Refaeilzadeh2016} to various learning rates, message-passing steps and zone of influence radii.
    The first step of the cross-validation process involved training a GNN framework for 5 epochs for each of the tested parameters.
    After training was completed, 12 cases from the validation set (each involving up to 100 time-steps) were randomly chosen to obtain the network's performance.   
    The performance was computed using the averaged maximum percent errors in the predicted crack lengths across the 12 validation cases. 
    The resulting averaged length percent errors for the learning rates, the message-passing steps and the zone of influence radii are shown in Figure \ref{fig:cross_validation}. 
    
    Figure \ref{fig:LR_Valid} shows the cross-validation results for learning rates of {$5\times 10^{-5}$, $1\times 10^{-4}$, $5\times 10^{-4}$, and $5 \times 10^{-3}$ } (gray) against our model’s learning rate of {$1\times 10^{-3}$} (red).
    The model with the lowest percent error was for learning rate of {$1\times 10^{-3}$} at $3.09 \pm 0.72\%$, compared to {the smaller learning rate of $1\times 10^{-4}$ with} error of $3.50 \pm 0.85 \%$.
    {Therefore, we chose the optimal learning rate of $1\times 10^{-3}$ for our model.}
    
    \begin{figure}
        \centering 
        \begin{subfigure}[c]{0.32\textwidth}
            \centering
            \includegraphics[width=0.94\linewidth]{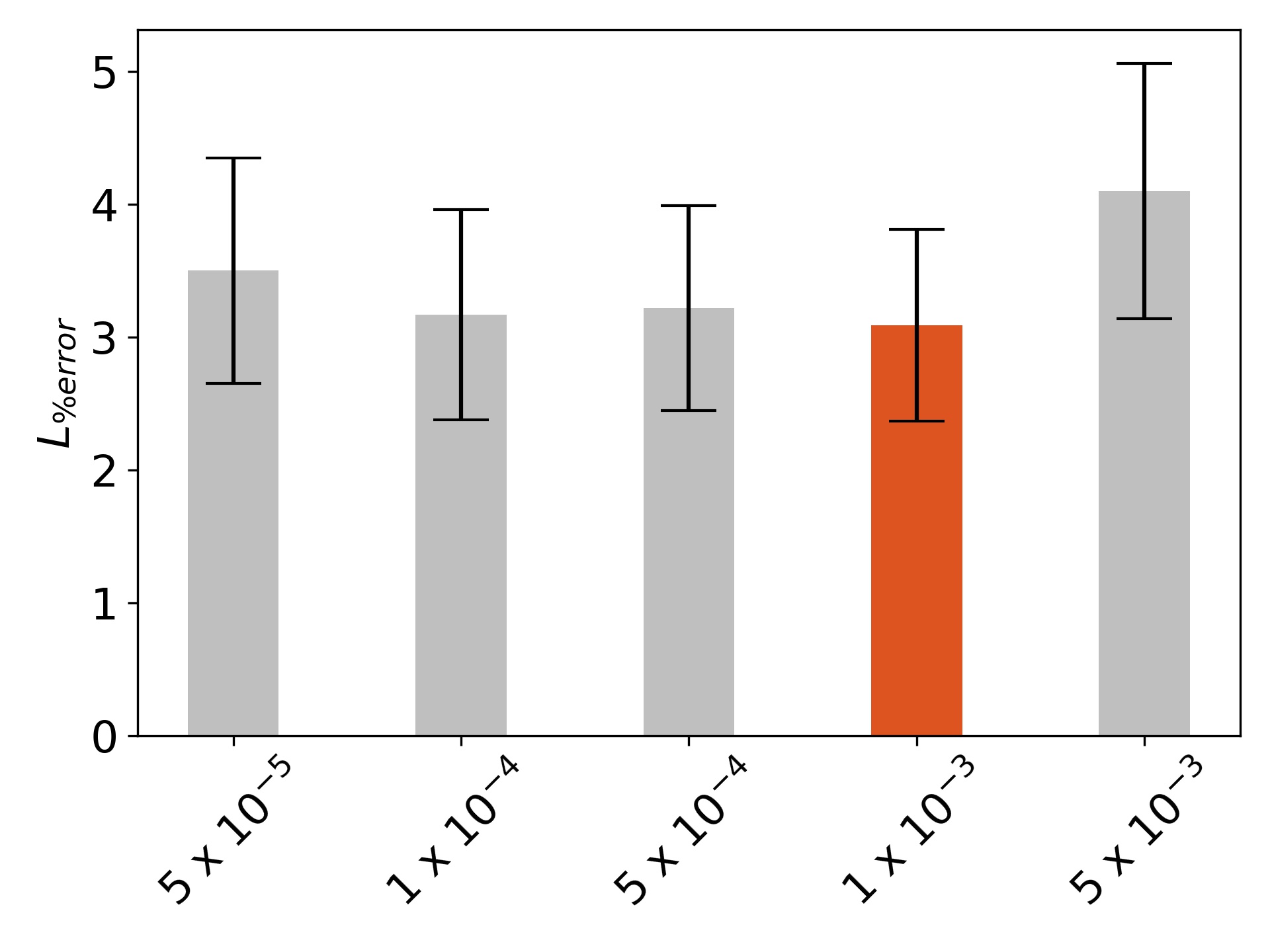}
            \centering
            \caption{Learning rates}
            \label{fig:LR_Valid}
        \end{subfigure}
        \centering
        \begin{subfigure}[c]{0.32\textwidth}
            \centering
            \includegraphics[width=0.98\linewidth]{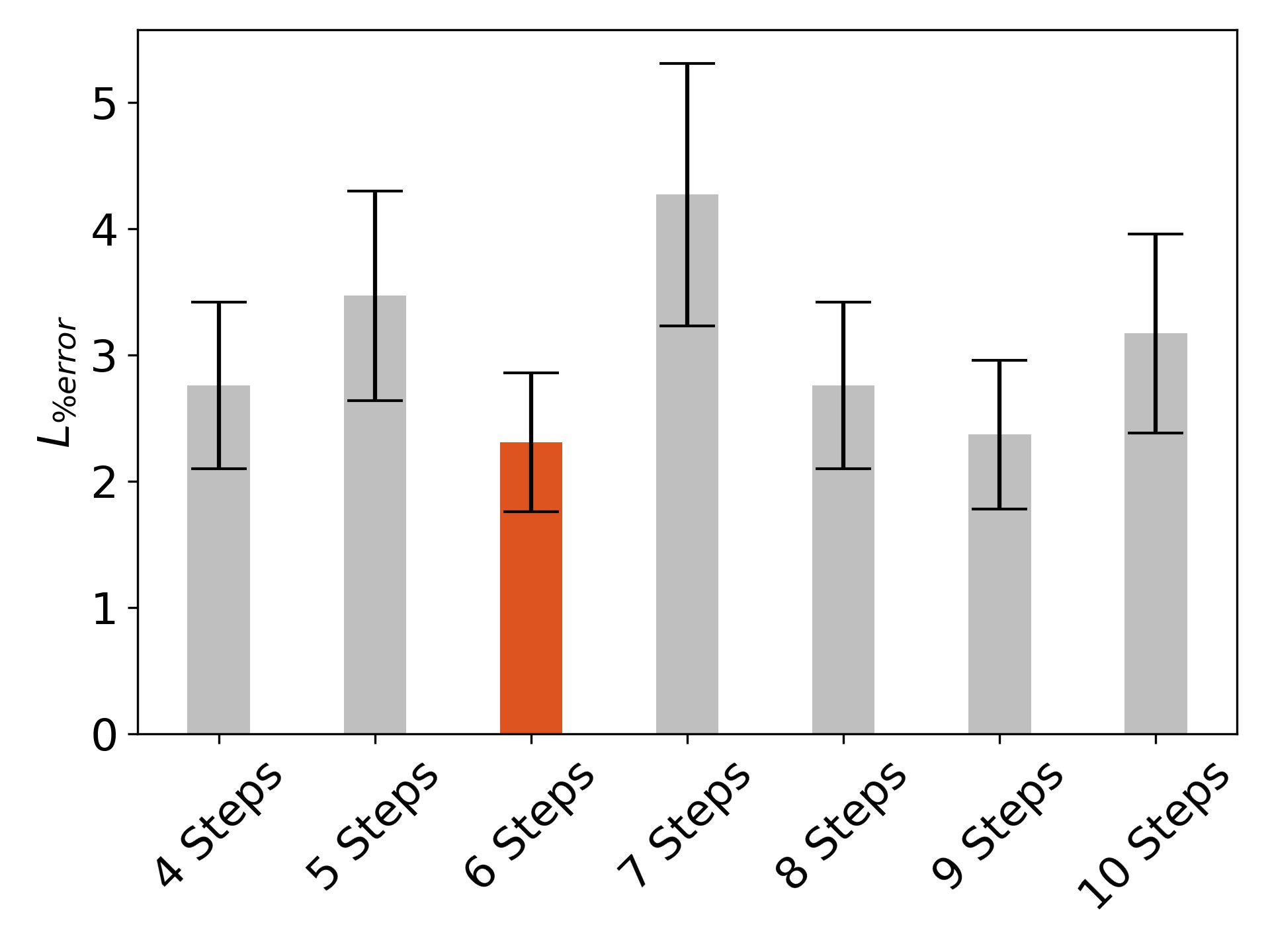}
            \centering
            \caption{Message-passing steps}
            \label{fig:MSteps_Valid}
        \end{subfigure}
        \centering
        \begin{subfigure}[c]{0.32\textwidth}
            \centering
            \includegraphics[width=0.98\linewidth]{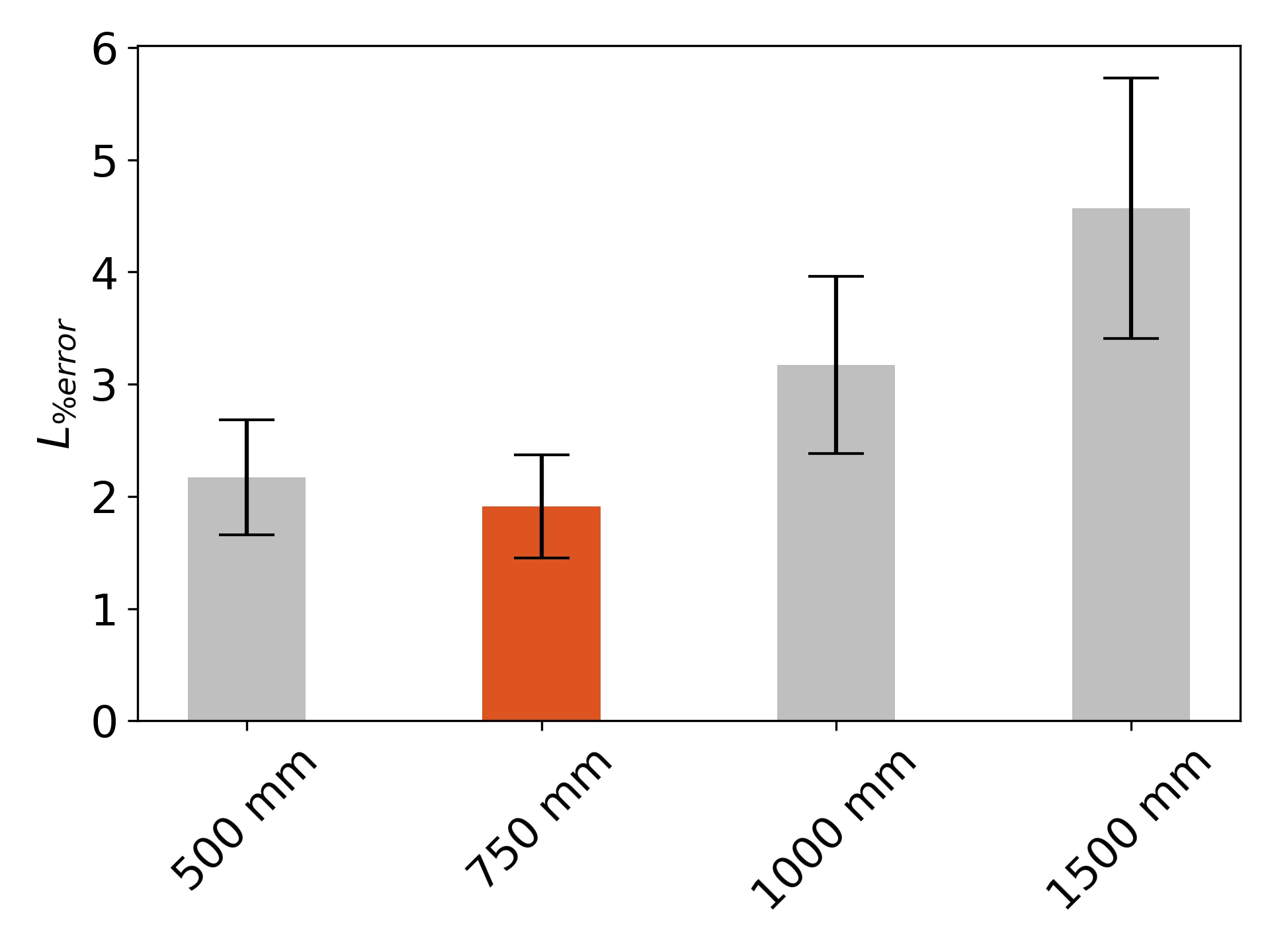}
            \centering
            \caption{Zone of influence radii}
            \label{fig:RC_Valid}
        \end{subfigure}
        \caption{Cross-validation results for a) learning rates $5 \times 10^{-5}$, $1 \times 10^{-4}$, $5 \times 10^{-4}$, and $5 \times 10^{-3}$ shown in gray, and our model's learning rate $1 \times 10^{-3}$ shown in red, b) message-passing steps of 4, 5, 7, 8, 9, and 10 shown in gray, and our model's message-passing steps of 6 shown in red, and c) zone of influence radii 500mm, 1000mm, and 1500mm shown in gray, and our model's zone of influence radius of 750mm shown in red.}
        \label{fig:cross_validation}
    \end{figure} 
    
    Figure \ref{fig:MSteps_Valid} shows the resultant averaged length percent errors for message-passing steps of {4,} 5, 7, 8, 9, and 10 (gray) against our model’s message-passing steps of {6} (red).
    The model with the lowest percent error was for message-passing steps of $M = 6$ at $2.31 \pm 0.55 \%$, compared to {the smallest number of message-passing steps of $M = 4$ with} error of $2.76 \pm 0.66 \%$.
    Similar to the cross-validation results for the learning rates, the optimal message-passing steps parameter of 6 {was} used in {this} work to further optimize our GNN model.
    However, we note that a higher number of message-passing steps (e.g., $M=10$) may increase training and simulation times \cite{pfaff2021learning}, thus, requiring multiple workers across GPUs to achieve similar speed-up \cite{zhang2021gmlp}.
    
    Finally, we tested the zone of influence radius; increasing the zone of influence increases the number of neighbors (number of nodes' and edges' relations) for each crack-tip.  
    Figure \ref{fig:RC_Valid} shows the resultant averaged length percent errors for zone of influence radii 500mm, { 1000mm}, and 1500mm (gray) against our model’s zone of influence of { 750mm} (red).
    We observe from Figure \ref{fig:RC_Valid} that a smaller zone of influence of 750mm achieves the least error at $1.91 \pm 0.46 \%$, compared to {a larger zone of influence of 1000mm with} error of $3.17 \pm 0.79 \%$.
    By further reducing the zone of influence to 500mm the error then increases to $2.17 \pm 0.51 \%$.
    We also note that the highest percent error at $4.57 \pm 1.16 \%$ was obtained for 1500mm.
    This higher error may be due to the fact that a large connectivity radius may lead to redundant connections (edges) between far-away crack-tips which do not influence each other significantly.
    In other words, additional connections between far-away crack-tips may oversample high-resolution areas in cases where propagating crack-tips are particularly influenced by their closer neighbors \cite{li2021learning}. }

\section{Results}\label{sec:Results}

    As explained in Section \ref{subsect:dataset} the testing dataset involved {15} simulations for each case of varying number of microcracks, resulting in a total of {225} simulations used in the error computations.
    We analyzed the framework's ability to predict microcrack interaction, propagation, and microcrack length growth. 
    We then performed error analyses on the predicted length growth, {predicted final crack path} {and} effective stress intensity factors $K_{eff}$,{ and performance comparison of the Microcrack-GNN with two additional baseline networks}.
    Lastly, we tested the computational time of Microcrack-GNN against XFEM for increasing number of microcracks in the domain. 
    In this section, we present the results corresponding to each analysis.
    Lastly, we tested the computational time of Microcrack-GNN against XFEM for increasing number of microcracks in the domain. 
    In this section, we present the results corresponding to each analysis.

    \subsection{Prediction of microcrack interaction, propagation, and coalescence}\label{sect:crack_growth} 
        
        \begin{figure}
            \includegraphics[width=\linewidth]{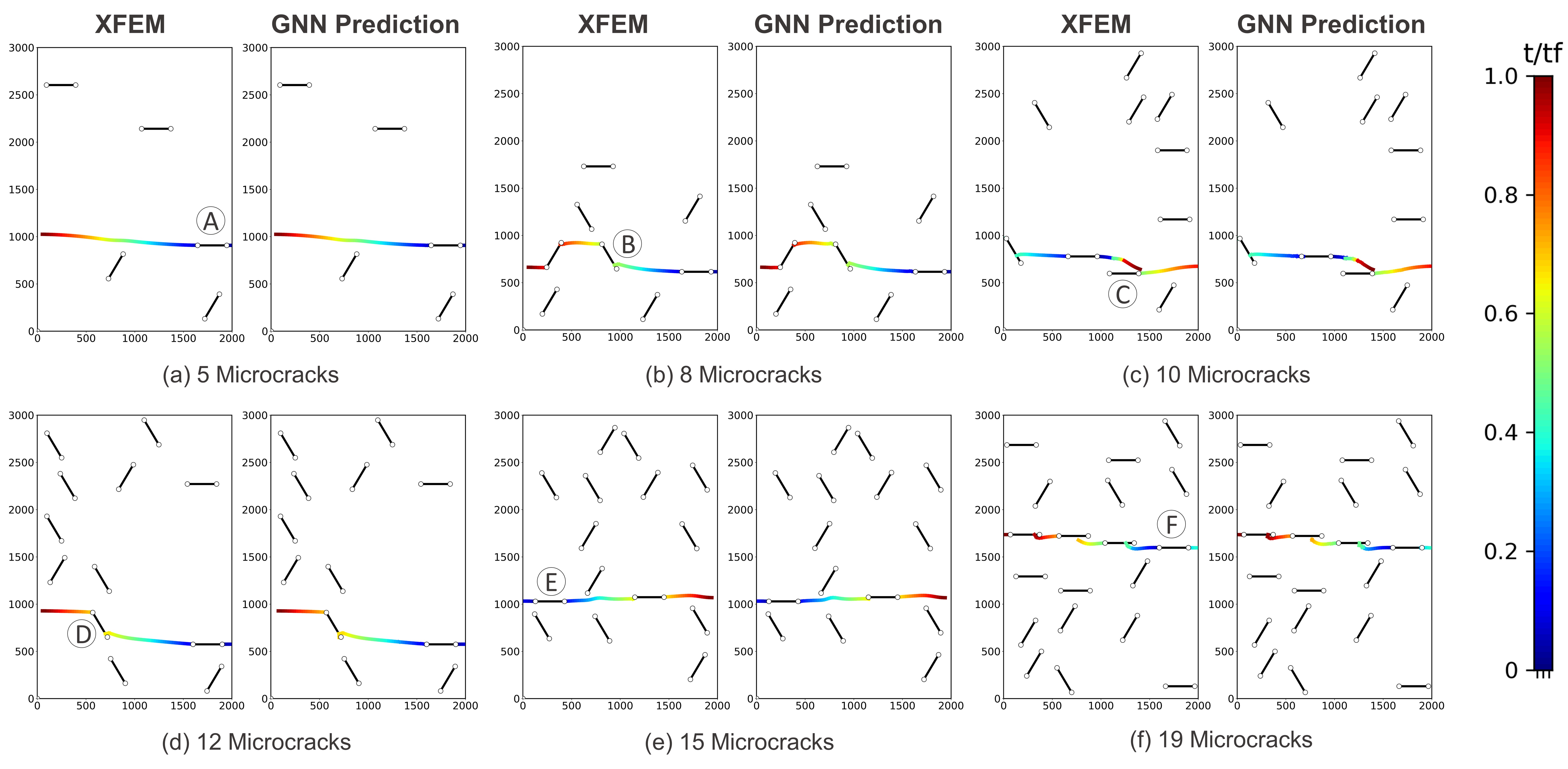}
            \caption{XFEM simulations versus GNN prediction of crack propagation and coalescence, for cases involving (a) 5, (b) 8, (c) 10, (d) 12, (e) 15, and (f) 19 microcracks. 
            The crack paths are colored based on $t/t_f$ where $t_f$ is the final simulation time for a given case.}
            \label{fig:emulation_5to19}
        \end{figure}
        To analyze the framework's prediction of microcrack interaction, propagation, and coalescence, we present predictions for 6 problem classes, each class involving a different number of microcracks, i.e., $C=5,8,10,12,15,$ and $19$ 
        from the XFEM dataset.
        Figure \ref{fig:emulation_5to19} shows crack evolution and coalescence with time as predicted by the \text{Microcrack-GNN} and compared to the XFEM simulations (see supplementary material for simulation videos).
        For $C=5$ (Figure \ref{fig:emulation_5to19}a), where only one crack propagated through the domain, the predicted crack path was nearly identical to the ground truth. 
        Additionally, for $C=12$ (Figure \ref{fig:emulation_5to19}d) and $C=15$ (Figure \ref{fig:emulation_5to19}e), two microcracks propagated and coalesced at approximately $60\%$ of the final time to failure.
        Comparing the XFEM predictions to the {Microcrack-GNN} predictions, both cases show nearly identical paths, depicting the framework's ability to predict cases involving coalescence of two microcracks.
        Additionally, two slightly more complex cases are shown in Figures  \ref{fig:emulation_5to19}b and \ref{fig:emulation_5to19}c for $C=8$ and $10$, respectively, where three microcracks coalescence through the domain.
        For the case of $C=8$ shown in Figure \ref{fig:emulation_5to19}b, the predicted crack paths and the coalescence period are virtually indistinguishable compared to the XFEM model.
        Similar to the previously mentioned cases, for $C=8$ the GNN framework was able to simulate coalescence of three microcracks with high accuracy.
        However, a more complex fracture path was obtained for $C=10$ (Figure \ref{fig:emulation_5to19}c) where errors in the predicted crack paths are evident.
        For instance, at approximately $t=0.4 t_f$ two cracks coalesced (left-most and middle crack), which caused a switch in the propagating crack-tip.
        During this switch for the propagating crack-tip a slight discontinuity in the predicted crack-tip is seen, overlapping the existing crack path (i.e., predicting crack growth in the opposite direction).
           
        A more intricate fracture scenario $C=19$ is presented in Figure \ref{fig:emulation_5to19}f.
        In this simulation the number of interacting microcracks increased to four; starting from the right-most microcrack denoted by ``F'', and following a sequentially coalescence until the left-most microcrack.
        In this case, we observe small deviations between {Microcrack-GNN} and XFEM results.
        For example, at the locations of crack coalescence (approximately $t=0.4t_f$ and $0.9t_f$ for the first and second crack coalescence period, respectively) the prediction shows a slight kink in the crack path (i.e., rough crack path during coalescence period). 
        However, taking these small errors into consideration, throughout the rest of the simulation the predictions follow a similar sequential propagation trend and failure path, approximately indistinguishable to the XFEM model.
        As a result, these findings confirm the developed {Microcrack-GNN} framework's capability to accurately simulate crack propagation and coalescence for higher-complexity cases involving up to $C = 19$ microcracks.

    \subsection{Microcrack length growth}\label{sect:crack_length}  
      
        \begin{figure}
            \centering 
            \includegraphics[width=\linewidth]{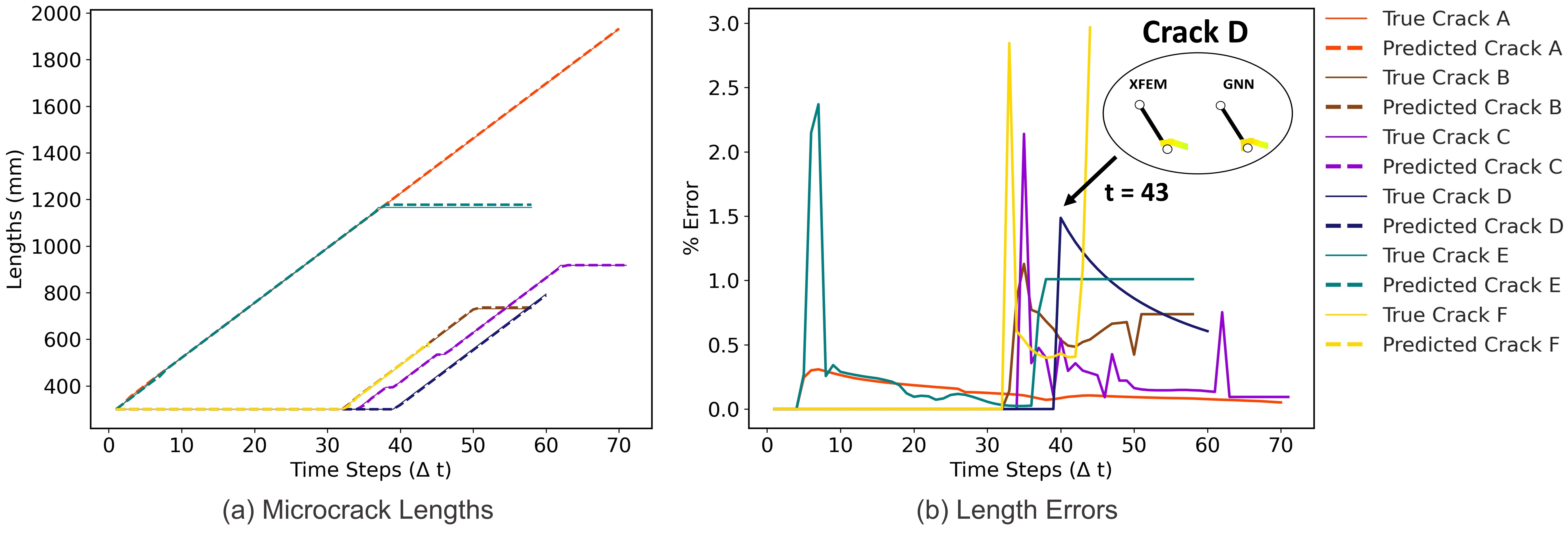}
            \centering
            \caption{The evolution of (a) crack length vs time for high-fidelity simulations and Microcrack-GNN predictions for cracks $A, B, \ldots, F$ labeled in Figure \ref{fig:emulation_5to19}. (b) Relative error between Microcrack-GNN and XFEM predicted crack lengths.}
            \label{fig:length_vs_time}
        \end{figure}
        
        Next we compare crack lengths as a function of time for another quantitative verification of the Microcrack-GNN's ability to accurately predict crack growth and coalescence.
        For this analysis, we computed the crack length growth as a function of time as shown in Figure \ref{fig:length_vs_time}, where for each simulation case, a propagating microcrack was used to track the change in length as depicted in Figure \ref{fig:emulation_5to19} by {A}, {B}, {C}, {D}, {E}, and {F}.
        For microcrack {A}, we observe a linear increase in the length starting at the initial length of $300 \ mm$ and reaching approximately $1900 \ mm$.  
        Comparing the XFEM and Microcrack-GNN predicted length from Figure \ref{fig:emulation_5to19}a and Figure \ref{fig:length_vs_time}, the predicted crack length is approximately identical to the ground truth.
        Furthermore, for the cases involving 8 and 10 microcracks where three microcracks coalesce, a similar high accuracy in the length prediction was also obtained {with intermediate jumps in error during crack coalescence}. 
        As shown in Figure \ref{fig:length_vs_time}, both the XFEM and the Microcrack-GNN predicted crack lengths for cracks B and C are seen to overlap throughout the complete simulation.
        
        For crack {D} (for $C=12$), the crack length is seen to remain at the initial length of $300 \ mm$ throughout most of the simulation, until reaching approximately time-step 43.
        At this time-step, crack {D} connects with the already-propagating microcrack (right-most crack) and we see a linear increase in length with time, along with {a spike of approximately 1.5\% in relative error}.
        {To understand the source of this spike in relative error, the zoomed-in region shown in Figure \ref{fig:length_vs_time}b depicts the time at which Crack D shown in Figure \ref{fig:emulation_5to19}d connects with the already-propagating microcrack for both the XFEM and GNN models.
        While the XFEM model generates a smooth connection between both cracks, the GNN model depicts a jump downwards once coalescence has occurred. 
        Thus, this downward jump creates an additional spike in error for crack D, whereby} {comparing} the Microcrack-GNN predicted length to XFEM, we note that {Microcrack-GNN} predicts final longer crack length than XFEM ({approximately $0.55\%$} relative error at $t=t_f$).
        For crack E ($C=15$), the length increases linearly until reaching a length of approximately $1200 \ mm$ during time-step 38. 
        Similar to crack D, the Microcrack-GNN predicted final crack length of crack E is slightly higher than the XFEM ({$0.9\%$} relative error at $t=t_f$).
        These results obtained for $C=12$ and {$C=15$} imply that the {Microcrack-GNN} framework may be predicting a slightly faster crack growth (i.e., simulation) compared to XFEM.
        
        A similar analysis for the more complex case of $C=19$, initially the right-most crack ({F}) is first to propagate, following a linear increase in the crack length as shown in Figure \ref{fig:length_vs_time}.
        During this phase, the Microcrack-GNN predicted crack length is nearly identical to XFEM predictions, until reaching approximately time-step 13 where we see a slight deviation between predictions. 
        Comparing the growth in length for crack {F} to the crack paths from Figure \ref{fig:emulation_5to19}f, we may conclude that the resulting small difference in the predicted length is linked to crack path deviations at the points of crack coalescence as mentioned previously.
        The findings shown in Figure \ref{fig:length_vs_time} for cracks A - F suggests that the accuracy in the predicted crack paths, and crack lengths may not depend on the initial number of microcracks in the system.
        Thus, indicating that deviations in the predictions may be influenced by the initial configuration of the system itself (e.g., cracks position and orientation).

    \subsection{Errors in final crack path}\label{sec:crack_path_error}
        
\begin{figure}
            \centering 
            \includegraphics[width=0.75\linewidth]{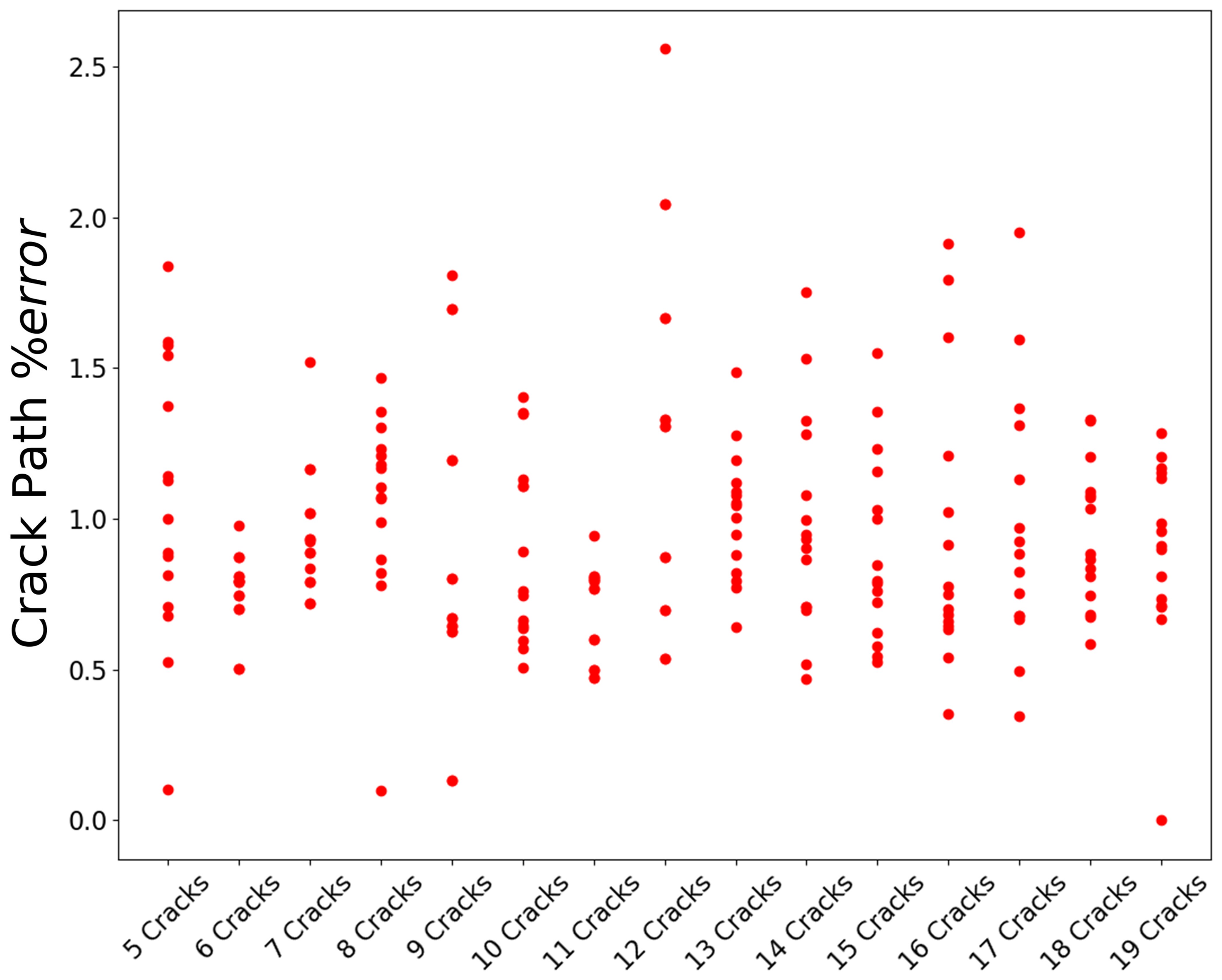}
            \centering
            \caption{Maximum percentage errors for crack path predictions for all test cases (225 test cases).}
            \label{fig:spline_plot}
        \end{figure}

        {In Section \ref{sect:crack_length}, we presented a time deviance error analysis for crack length where we computed the maximum error between the XFEM-based crack length and the predicted crack length at any given time.
        We note that the Microcrack-GNN framework may predict faster or slower crack propagation in some cases.
        However, the final predicted crack path qualitatively was nearly identical to the XFEM crack path as shown in Figure \ref{fig:emulation_5to19}. 
        In this section we also perform a spatial deviance analysis to characterize the errors in the final predicted crack paths.
        For this analysis, we computed the maximum percent error in distance between the XFEM and Microcrack-GNN crack paths.
        
        Figure \ref{fig:spline_plot} depicts the computed crack path errors for the entire test dataset (225 cases).
        We note that that compared to time-wise crack length errors (Figure \ref{fig:Length_testset_error} in Appendix), the errors for the predicted final crack paths are significantly lower.
        The maximum error was obtained for the case of 12 microcracks at approximately $2.53\%$, while the minimum error was obtained for the case of 19 microcracks at approximately $0.004 \%$.
        As a result, while the time deviance errors for the predicted crack lengths showed higher errors, Figure \ref{fig:spline_plot} shows that the Microcrack-GNN framework predicted final crack paths with good accuracy (as low as 2.53\% for the test set)} 
        
    \subsection{Errors on effective stress intensity factor}\label{subsec:results_Keff}

        To quantify errors in prediction of the stress intensity factors, we present 6 cases of varying number of microcracks ($C=5,8,10,12$ and $19$). 
        First, the effective stress intensity factors were computed using the Mode-I and Mode-II stress intensity factors as
        \begin{flalign}
            {K_{eff}}  = \sqrt{(K_{I})^{2} + (K_{II})^{2}}.
        \label{eq:Keff}
        \end{flalign}
        
        \begin{figure}
            \centering 
            \includegraphics[width=0.75\linewidth]{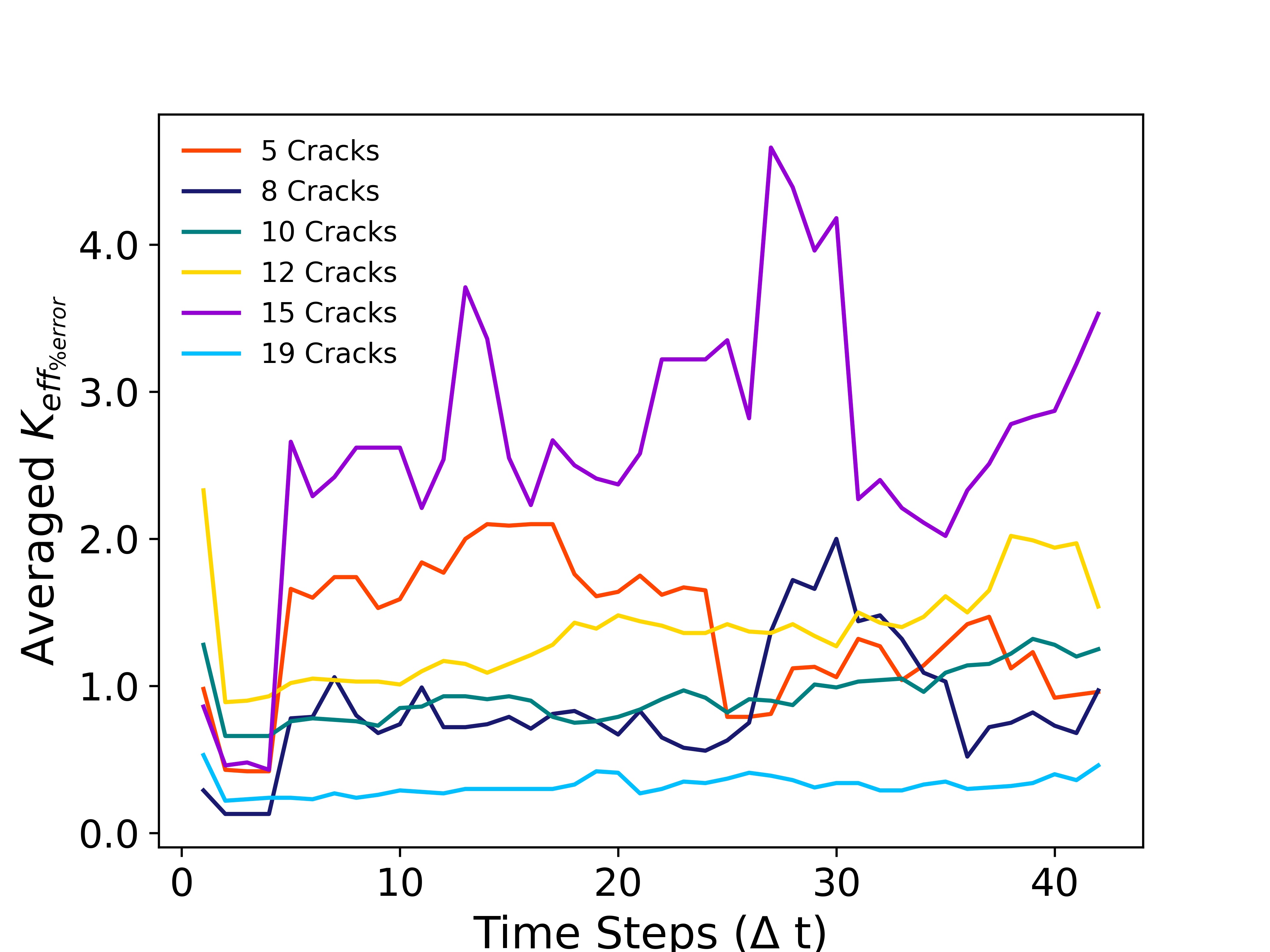}
            \centering
            \caption{``Averaged'' maximum percentage errors for the effective stress intensity factors for cases involving 5, 8, 10, 12, 15, and 19 microcracks as a function of time.}
            \label{fig:Keff_error}
        \end{figure}

        For error calculations, we considered the crack-tips where $K_{eff} \geq K_{crt}$.
        This is because these cracks are most likely to propagate at any given time-step.
        For each simulation and each time-step we compute the error in $K_{eff}$ as
        \begin{flalign}
            && {K_{eff}}_{\% error} = \max_{s\in N_{crt}^t} \left( \frac{\left| {K_{eff}}_{\ Pred}^{t} - {K_{eff}}_{\ True}^{t} \right|}{{K_{eff}}_{\ True}^{t}} \right)_{s} \times 100 &&  \{t = 1, 2, \dots ,  T_{f}\},
        \label{eq:Keff_error}
        \end{flalign}
        where $N_{crt}^t$ is the number of cracks tips with $K_{eff}\geq K_{crt}$ at any given time $t$, ${K_{eff}}^t_{Pred}$ is the predicted $K_{eff}$ at time $t$ by the Microcrack-GNN, and ${K_{eff}}^t_{True}$ is the true $K_{eff}$ at time $t$ by the XFEM framework.
        
        For each $C$, the test-set contained 10 simulations. 
        Figure \ref{fig:Keff_error} shows the average of $ {K_{eff}}_{\% error}$ for 10 simulations for $C=5,8,10,12,15,\text{ and } 19$ as a function of time $t$.
        From Figure \ref{fig:Keff_error}, it can be observed that the highest percentage errors from all cases of varying number of microcracks were obtained for the case involving 15 microcracks reaching approximately $4.80\%$. 
        Additionally, the minimum percentage error was obtained for the case of 19 microcracks at approximately $0.20\%$.  
        We note that the percent errors did not increase as a function of the number of initial microcracks.
        For instance, for the cases of 5, 8, 10, 12, and 15 microcracks the resultant errors were higher compared to the case of 19 microcracks.   
        These results suggest that the errors in the predicted stress relations are likely determined by the initial orientations and positions of the interacting microcracks, rather than to the complexity of the problem (i.e., number of initial microcracks).

    \subsection{Additional Baselines}

    {To compare the performance of the developed Microcrack-GNN to other ML models, we developed and trained two additional models. 
    The training consisted of 5 epochs, similar to the Microcrack-GNN using two loss functions, Mean-Squared Error (MSE) and the Mean Absolute Error, also known as L1 loss. }

    \begin{enumerate}
        \item {RCNN: The first baseline model involved a Recurrent Convolutional Neural Network (RCNN) with two identity convolution layers, two batch normalization layers followed by the ReLU activation function, and a Linear layer as the final output. 
        The input to the RCNN consisted of a $(4 \times 38 \times 6)$ matrix, where $4$ indicates the number channels (time sequence $\hat{\mathbf{T}}$), $38$ indicates the number of nodes (crack-tips), and $6$ indicates the number of features (crack orientation, x and y positions, propagating vs non-propagating crack-tips, and $K_I$ and $K_{II}$ stress intensity factors) for the time sequence $\hat{\mathbf{T}}$.}
        
        \item {REDNN: The second ML model involved a Recurrent Encoder-Decoder Neural Network (REDNN) with four convolution layers, one feed-forward layer, and 4 transpose convolutions; the ReLU activation function was used for each layer. 
        The input to the REDNN was consistent with the input used for the RCNN.}  
        
    \end{enumerate}

    {Once training was completed for each baseline model, we compared their performance to the Microcrack-GNN.
    For this, we computed the error in the predicted effective stress intensity factor, $K_{eff}$, and the error in the predicted crack length for the case of 12 microcracks shown in Figure \ref{fig:length_vs_time}.
    The resultant errors were then tabulated as shown in Table 1.
    From Table 1, we note that the RCNN model outperformed the REDNN model when predicting Mode-I and Mode-II stress intensity factors. 
    The RCNN model with L1 loss function resulted in a lower percent error of $12.71\%$ than the RCNN model with MSE loss function of $13.71\%$.
    Additionally, we note that the opposite performance is obtained in the predictions of crack length.
    The REDNN models significantly outperform the RCNN models when predicting crack length.
    The baseline model with the lowest error was the REDNN model with MSE loss function of $16.03\%$, compared to the REDNN model with L1 loss function of $17.09\%$.
    However, in both prediction cases, $K_{eff}$ and crack Length, the Microcrack-GNN significantly outperforms both the RCNN and REDNN models achieving the lowest percent error of $1.85\%$ for the effective stress intensity factor, $K_{eff}$, and $0.32\%$ for the crack length.
    Therefore, these results demonstrate the strength of the developed GNN compared to other popular ML baselines. }
    
    \begin{tabular}{||c c c||} 
        \hline
        \multicolumn{3}{|c|}{Table 1: Performance comparison of baseline models versus Microcrack-GNN} \\
        \hline
        Models& $K_{eff} \% Error$ &Length $\%$ Error\\
        \hline
        RCNN (L1)& $12.71 \%$ & $62.47 \%$ \\
        RCNN (MSE)&   $13.71 \%$  &$41.74 \%$ \\
        REDNN (L1)& $39.66 \%$ & $17.09 \%$ \\
        REDNN (MSE)& $35.94 \%$ & $16.03 \%$ \\
        Microcrack-GNN& $\textbf{1.85 \%}$ & $\textbf{0.32 \%}$ \\
        \hline
    \end{tabular}
    
    \subsection{Analysis time VS. number of microcracks} 
        
\begin{figure}
            \centering 
            \begin{subfigure}[c]{0.49\textwidth}
                \centering
                \includegraphics[width=\linewidth]{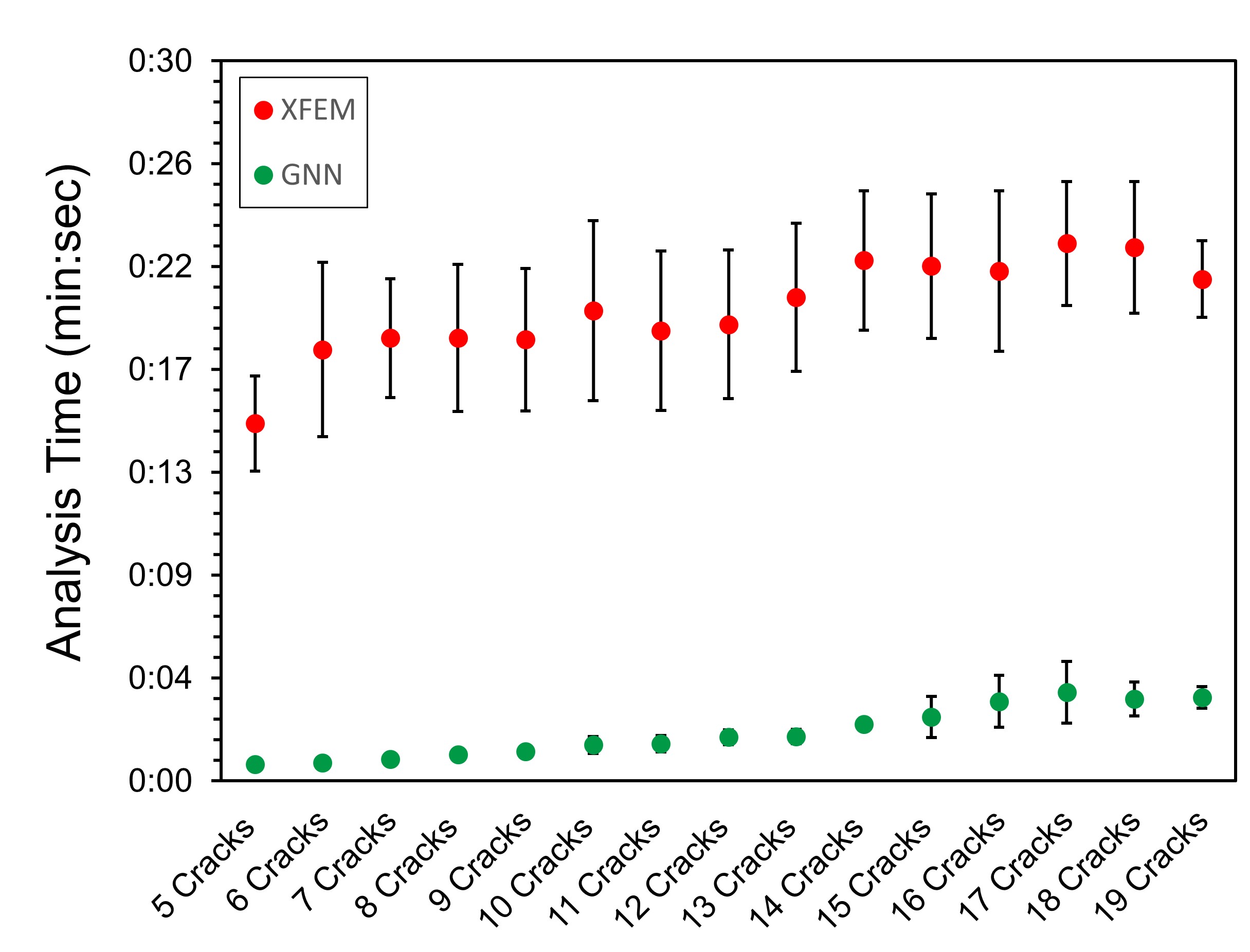}
                \centering
                \caption{GNN versus XFEM required CPU time}
                \label{fig:simulation_time}
            \end{subfigure}
            \centering
            \begin{subfigure}[c]{0.49\textwidth}
                \centering
                \includegraphics[width=\linewidth]{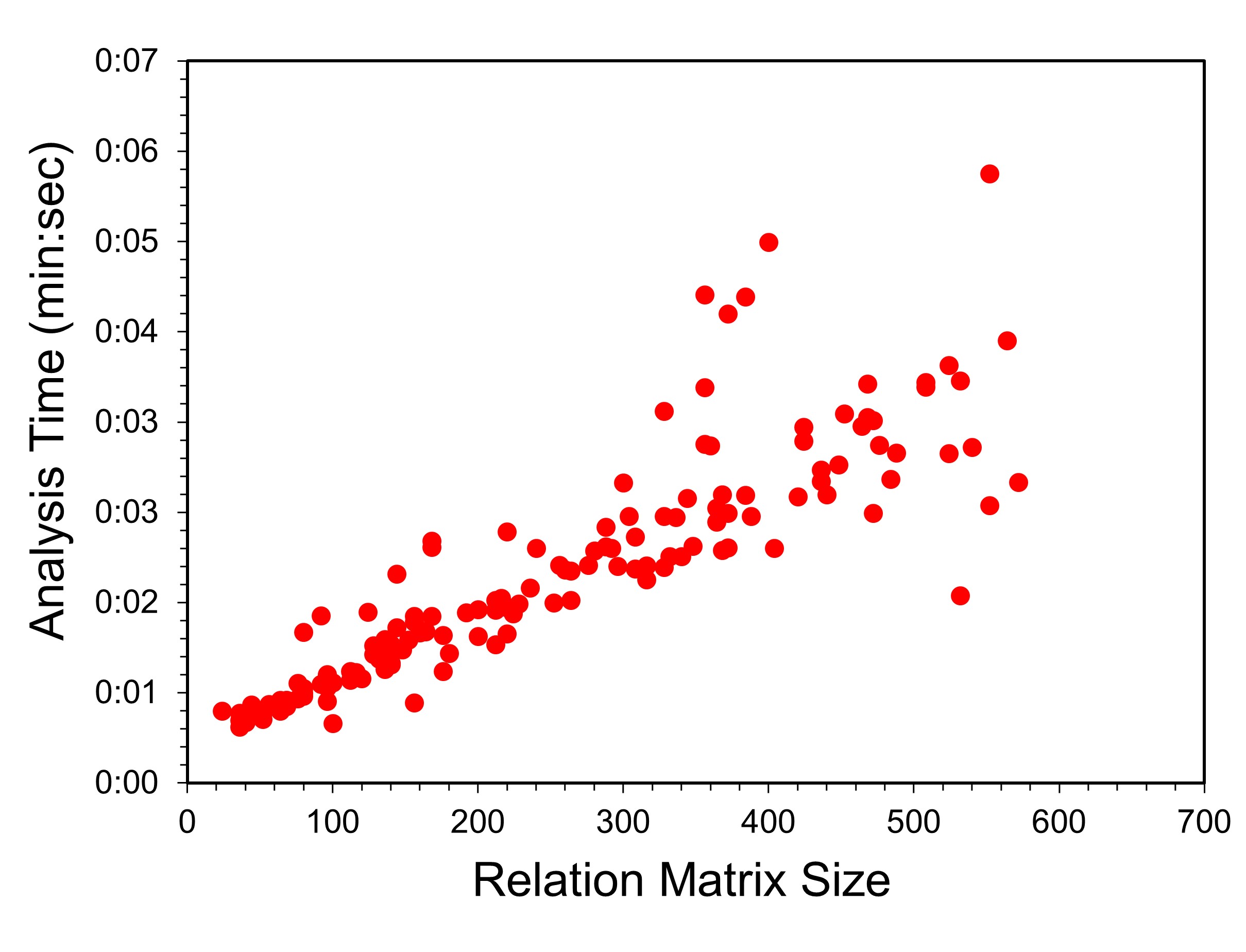}
                \centering
                \caption{Size of relation matrix VS. required CPU time}
                \label{fig:rel_time}
            \end{subfigure}
            \caption{Average required CPU time (min:sec) per simulation time frame versus a) number of initial microcracks (5 to 19) and b) size of relation matrix}
            \label{fig:time_analysis}
        \end{figure}

        Next, we test the computational time of Microcrack-GNN against XFEM for increasing number of microcracks in the domain.
        We compare the average CPU time needed per simulation time frame for Microcrack-GNN and XFEM for different $C$ values in Figure \ref{fig:simulation_time}.
        This analysis was performed {using an Intel(R) Core(TM) i3-10100 CPU @ 3.60GHz with 16Gb RAM} to test the framework's ability {to speed-up simulation time compared to XFEM. }
        We used a total of 10 simulations for each varying number of microcracks from the test-set in this analysis.
        From Figure \ref{fig:simulation_time}, we observe that the XFEM framework required significantly longer times compared to the Microcrack-GNN, {resulting in 6x to 25x speed-up when using the GNN framework}. 
        
        In this context, {both} the XFEM model {and the GNN framework show} an increasing trend in simulation times as number of initial microcracks increased.
        We note that Microcrack-GNN's computational performance is directly dependent on the number of initial microcracks {due to the increasing size of the relation matrix.
        In Figure \ref{fig:rel_time}, we show the required simulation time (min:sec) of the GNN framework as a function of the relation matrix size.  
        The relation matrix size depicts the number of edges in a given graph configuration, where a higher number of initial microcracks results in a higher number of edges in the graph.}
        Ultimately, Figure \ref{fig:simulation_time} shows a substantial {speed-up of up to 25x faster when using} GNN models for simulating higher-complexity fracture problems.
        As a future work, a larger range of initial microcracks may be simulated by the {Microcrack-GNN} framework with similar time performance to the cases shown in Figure \ref{fig:simulation_time}.

\section{Conclusion}\label{sec:Conclusion}

    To conclude, the integration of graph theory along with GNNs and fracture mechanics is a recent field of study which shows promise for speeding-up existing {high-fidelity fracture mechanics} models.
    The integration and development of such models {to simulate both crack propagation and stress evolution in brittle materials with varying number of initial microcracks have not been studied in previous work.}
    As shown in Figure \ref{fig:MicroCrack-GNN_structure}, we developed four GNNs capable of modeling the underlying physics for these problems.
    By integrating such GNNs, we obtained a 
    framework that can {dynamically} predict the future crack-tip positions and coalescence, crack-tip stress intensity factors, and the stress distribution throughout the domain {at each future time-step}.
    The results for the {$K_{I}$-GNN} and {$K_{II}$-GNN} showed high accuracies {for the test dataset} in the predicted $K_{eff}$ with a maximum relative error of $4.80\%$ compared to the XFEM method.
    The Microcrack-GNN framework also demonstrated high accuracy {for the test dataset} in the predicted crack lengths with maximum percent errors of $4.29\%$ and $1.01\%$.
    Additionally, Microcrack-GNN offers capability for simulating varying number of microcracks from 5 to 19, without additional modifications to any of the integrated GNN models.
    
    While the Microcrack-GNN framework predicts crack propagation and coalescence with good accuracy {on the test dataset}, we note various limitations in its current state.
    {Computations for Von Mises stresses throughout the domain are generated using LEFM superposition principle on the predicted Mode-I and Mode-II stress intensity factors which can lead to high errors ($> 8 \%$) in some cases.}
    The framework is not optimized and therefore has a long training time (5.18 hours on four NVidia T4 GPUs).
    We note that we used an in-house GNN implementation in Pytorch instead of existing libraries \cite{fey2019fast,tiezzi2020lagrangian}.
    The GNN back-end libraries can be optimized for performance with flexible GPU resource allocation, improved spatial message-passing and efficient parameterizations.
    The overall network architecture can also be optimized to reduce training time and preserve accuracy \cite{PERERA2021110524,kasim2020building,WILLJUICEIRUTHAYARAJAN20105775, hansen2016cma}.
    We also note that the XFEM-based model was not optimized to work in parallel using multiple CPUs. 
    These optimizations will be reserved for future work.
    
    With back-end and network optimizations, the {Microcrack-GNN} framework offers potential of further improvement to be a fast and accurate simulator for large number of cracks.
    The framework can also be extended to study ductile materials and even capture crack nucleation by utilizing dynamic graphs in future works.

\section{Acknowledgements}
Authors are grateful for the support of the Auburn University Easley Cluster for assistance with this work.
Financial support was also provided by the U.S. Department of Defense through the SMART scholarship Program (SMART ID: $2021-17978$).

\bibliographystyle{ieeetr}
\bibliography{library}

\appendix

\section{Test dataset Errors}

    As described in Section \ref{sect:Setup}, the test dataset involved  {15} simulations for each varying number of microcracks (5 to 19 microcracks) resulting in a total of {225} simulation.
    We emphasize that each simulation contains between 50-100 time-steps until failure, thus, resulting in a total test dataset size of {11,250 to 22,500} discreet time-steps. 
    Here we present crack length percent errors, effective stress intensity factors percent error, and Von Mises stress percent errors for all {225} simulations.
    
    \subsection{Length percent errors}
\begin{figure}
            \centering 
            \includegraphics[width=0.75\linewidth]{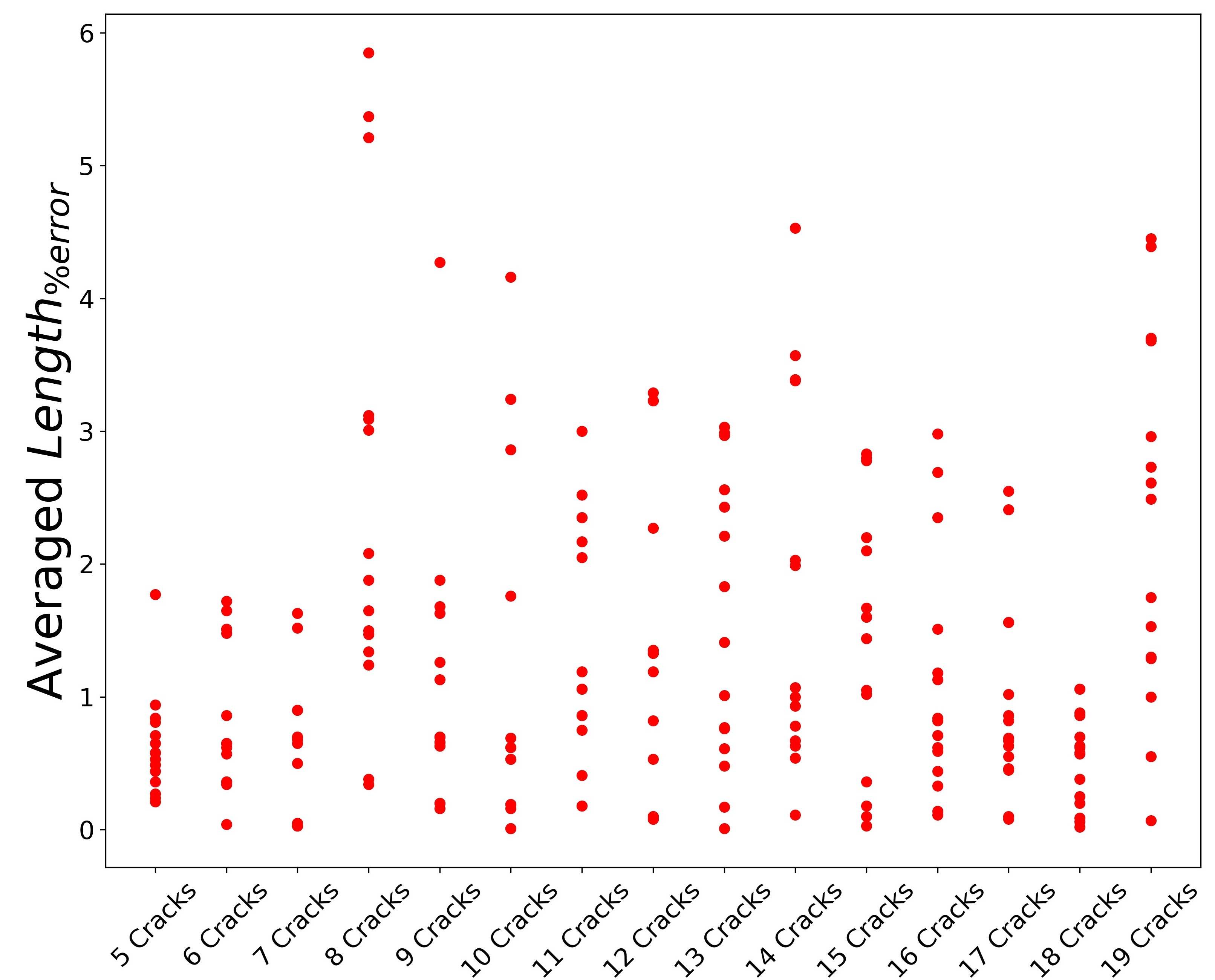}
            \centering
            \caption{Averaged maximum percentage errors of crack length predictions for all test cases (225 test cases).}
            \label{fig:Length_testset_error}
        \end{figure}

        For each test simulation, for each $C$, and for each time, we obtained the maximum percent error in crack.
        Next, for each test simulation, we computed the average across time resulting in 15 error points for each $C$, as shown in Figure \ref{fig:Length_testset_error}.  
        The resulting highest length error in the test dataset can be noted for test case {11} of 8 microcracks at approximately {$5.85\%$}  error.
    
        \begin{figure}
            \centering 
            \begin{subfigure}[c]{0.49\textwidth}
                \centering
                \includegraphics[width=\linewidth]{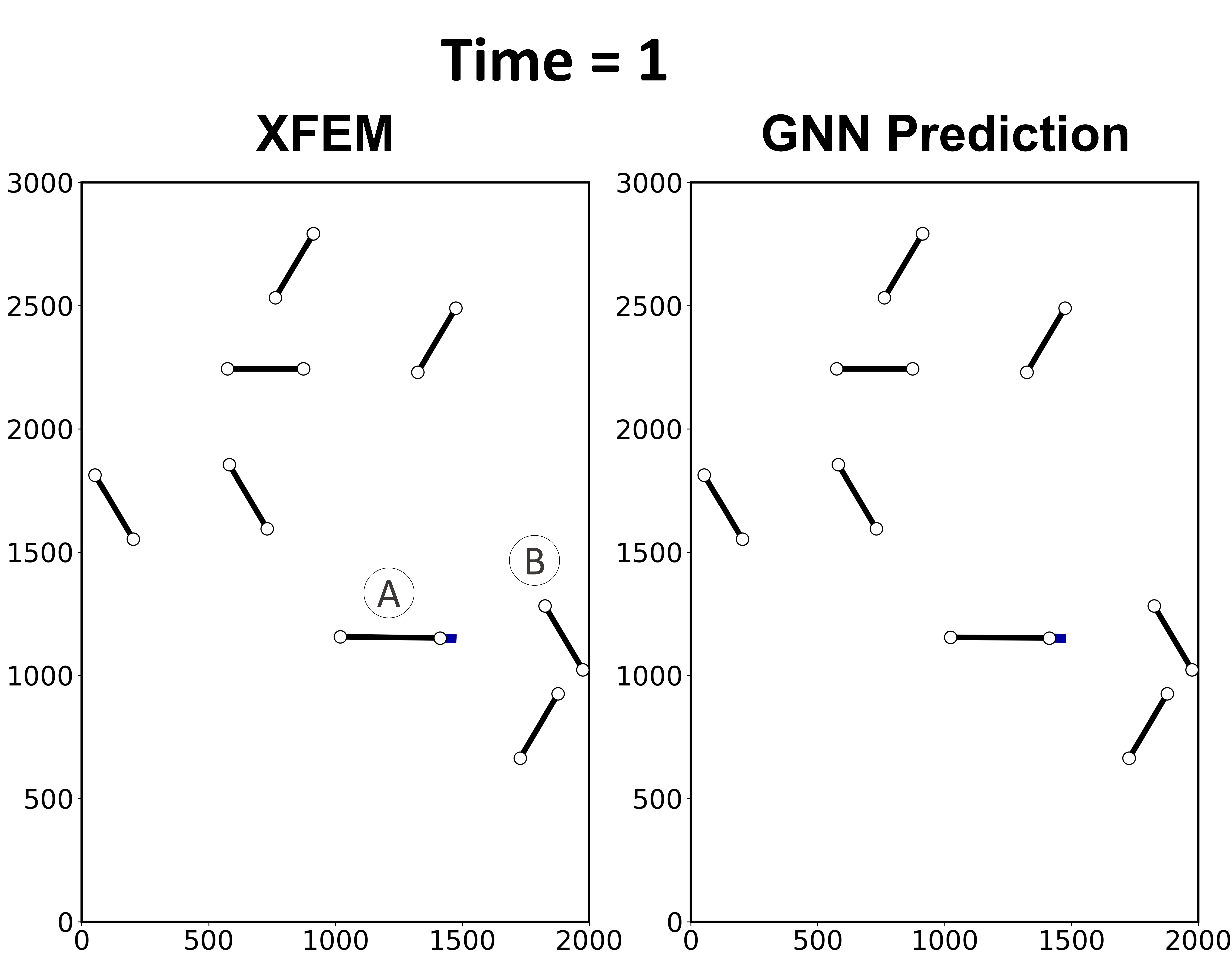}
                \centering
                \caption{Time-step 1 of 8 cracks}
                \label{fig:Time-step_2}
            \end{subfigure}
            \centering
            \begin{subfigure}[c]{0.49\textwidth}
                \centering
                \includegraphics[width=\linewidth]{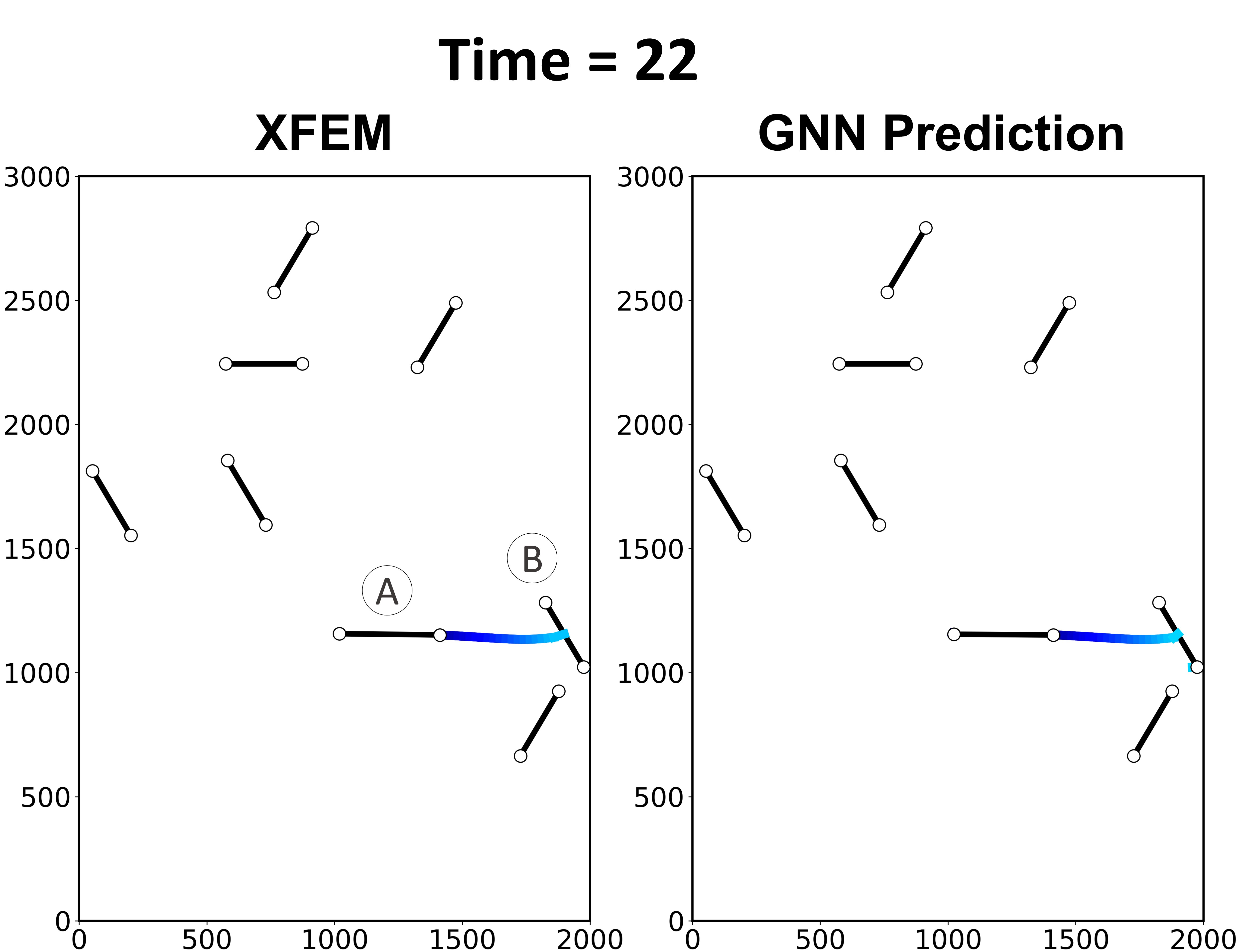}
                \centering
                \caption{Time-step 22 of 8 cracks}
                \label{fig:Time-step_22}
            \end{subfigure}
            \caption{Crack propagation for test case 11 of 8 microcracks at a) Time-step = 1 and b) Time-step = 22}
            \label{fig:CrackPath_8Cracks_Test11}
        \end{figure} 
        
        \begin{figure}
            \centering 
            \begin{subfigure}[c]{0.49\textwidth}
                \centering
                \includegraphics[width=\linewidth]{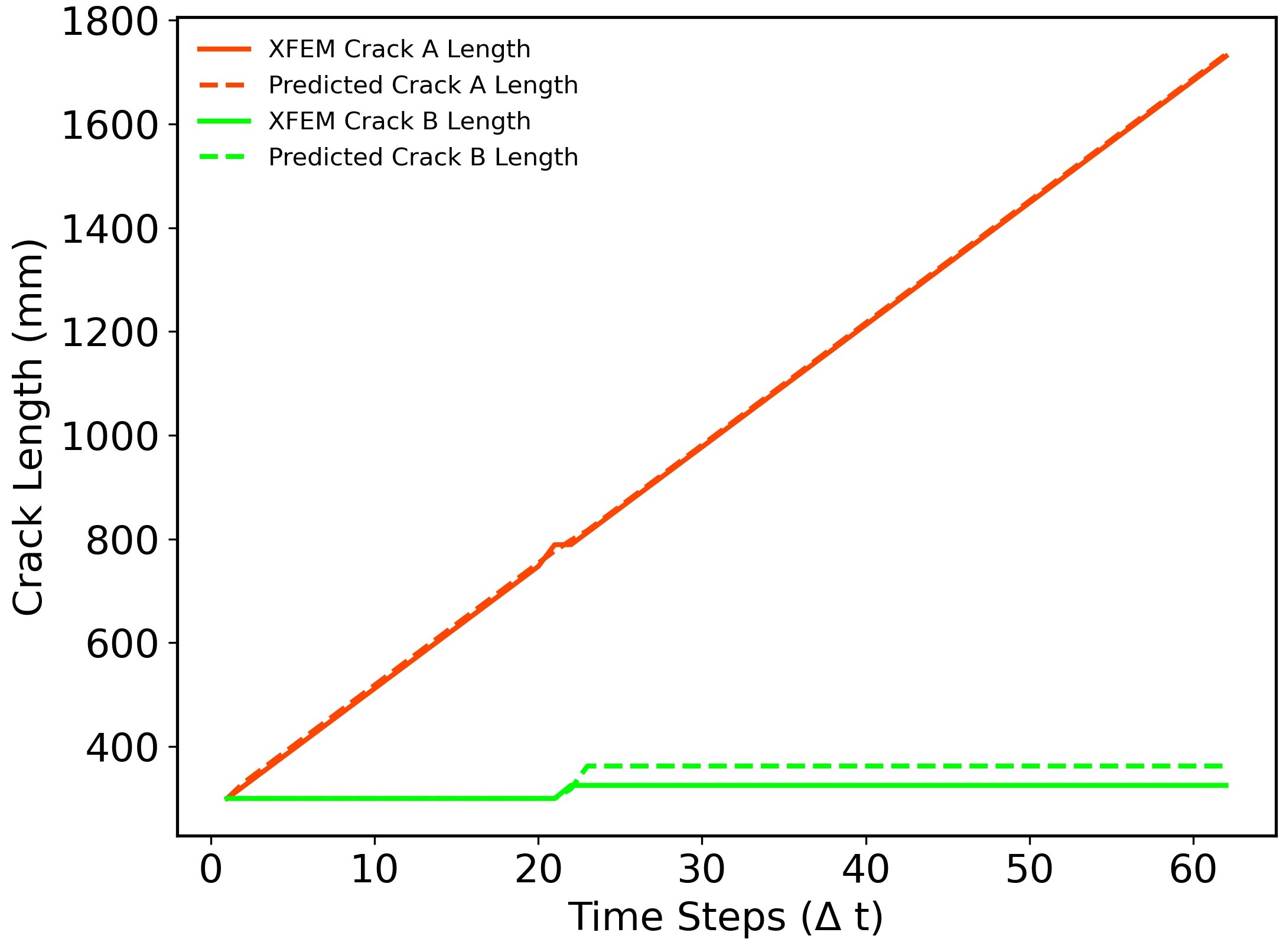}
                \centering
                \caption{Microcrack Length growth}
                \label{fig:LengthComparison_8Cracks_Test11}
            \end{subfigure}
            \centering
            \begin{subfigure}[c]{0.49\textwidth}
                \centering
                \includegraphics[width=\linewidth]{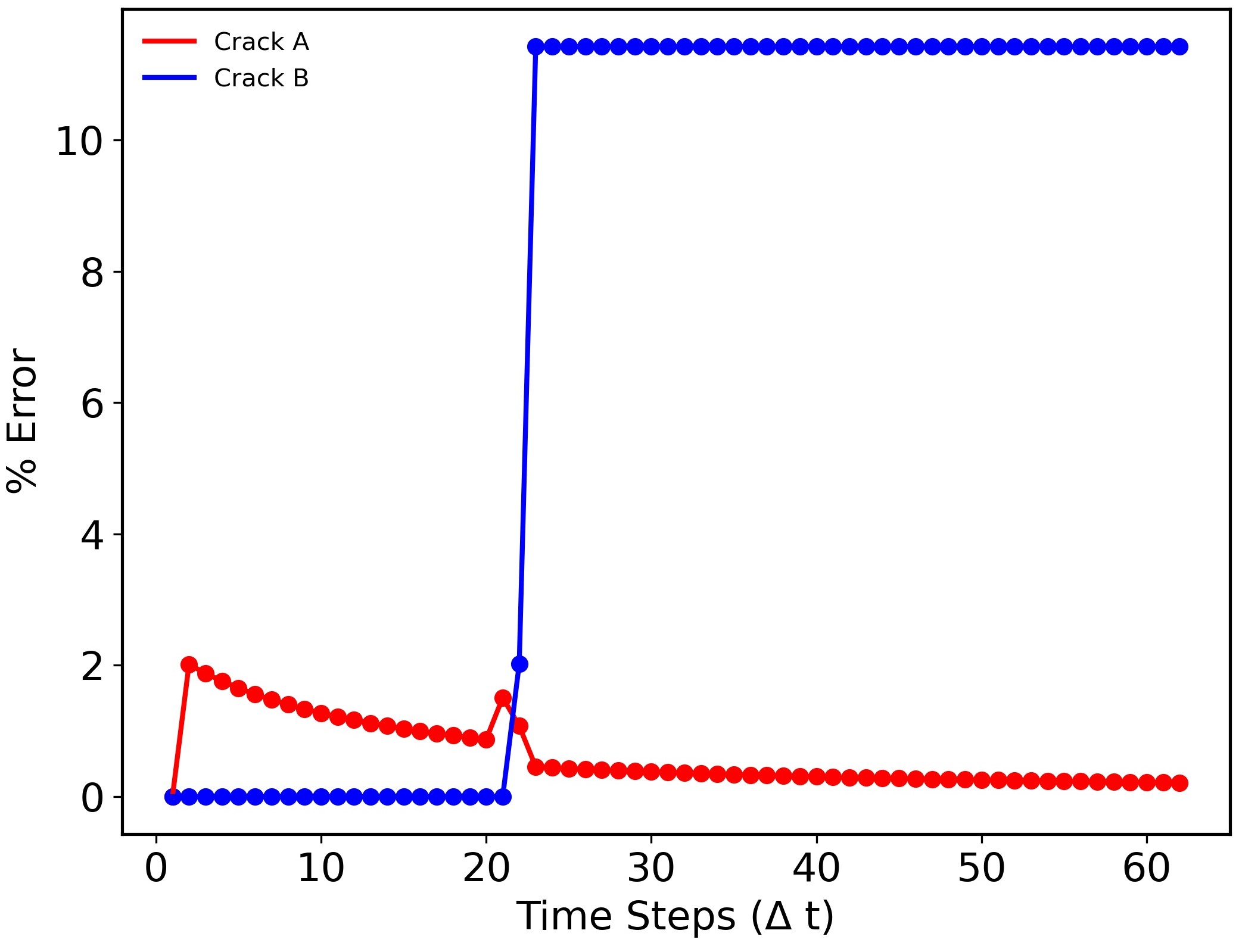}
                \centering
                \caption{Microcrack Length Error}
                \label{fig:LengthError_8Cracks_Test11}
            \end{subfigure}
            \caption{The evolution of a) crack length versus time for high-fidelity XFEM simulation and Microcrack-GNN simulation for test case 11 of 8 microcracks and b) Relative error between Microcrack-GNN and XFEM predicted crack lengths for test case 11 of 8 microcracks}
            \label{fig:8Cracks_Test11_Length}
        \end{figure} 
        
        To understand the source of error for test case {11} of 8 microcracks, we look at Figures \ref{fig:CrackPath_8Cracks_Test11} and \ref{fig:8Cracks_Test11_Length}.
        Figure \ref{fig:CrackPath_8Cracks_Test11} shows a qualitative comparison between the evolution of cracks A and B as predicted by XFEM and GNN at time-steps {1 and 22}. 
        Figure \ref{fig:LengthComparison_8Cracks_Test11} (and \ref{fig:LengthError_8Cracks_Test11}) shows the computed lengths of cracks A and B over time as predicted by XFEM and GNN (and errors).
        From \ref{fig:LengthComparison_8Cracks_Test11}, an initial jump in error of approximately {2.1\%} can be seen from Figure \ref{fig:8Cracks_Test11_Length} {for crack A} during time-step {1}. 
        Referring to Figure \ref{fig:Time-step_2}, the high jump in error originates from the predicted crack A's propagation being slightly directed towards the {negative y-direction to the right of the domain}, while the XFEM crack A's propagation follows a {straight} path towards the {right of the domain}. 
        This error results in a predicted crack length of {slightly} larger size compared to the XFEM-generated crack length, which can be also be seen from Figure \ref{fig:LengthComparison_8Cracks_Test11}.
        
        The highest jump in error occurs during time-step {22} also for crack {B} with approximately {10.8\%} error as shown in Figures \ref{fig:LengthComparison_8Cracks_Test11} and \ref{fig:CrackPath_8Cracks_Test11}.
        From Figures \ref{fig:Time-step_22} and \ref{fig:LengthComparison_8Cracks_Test11}, it can be seen at time-step {22} that crack {A} has already propagated during previous time-steps.
        {At time-step 22 the right-most crack tip of crack B propagates towards the right, thus, meeting the right edge of the domain}.
        {From Figure \ref{fig:Time-step_22}, at time-step 22} the XFEM-generated crack propagation direction is seen towards the {positive x-direction (right)}, and the Microcrack-GNN predicted crack propagation direction is towards the {negative x-direction (left)}.
        This incorrect prediction of crack path direction results in a larger predicted crack towards the remaining time-steps, and the largest percent error from the entire test dataset of {5.85\%} error.

    \subsection{Effective stress intensity factor percent errors}
    
\begin{figure}
            \centering 
            \includegraphics[width=0.75\linewidth]{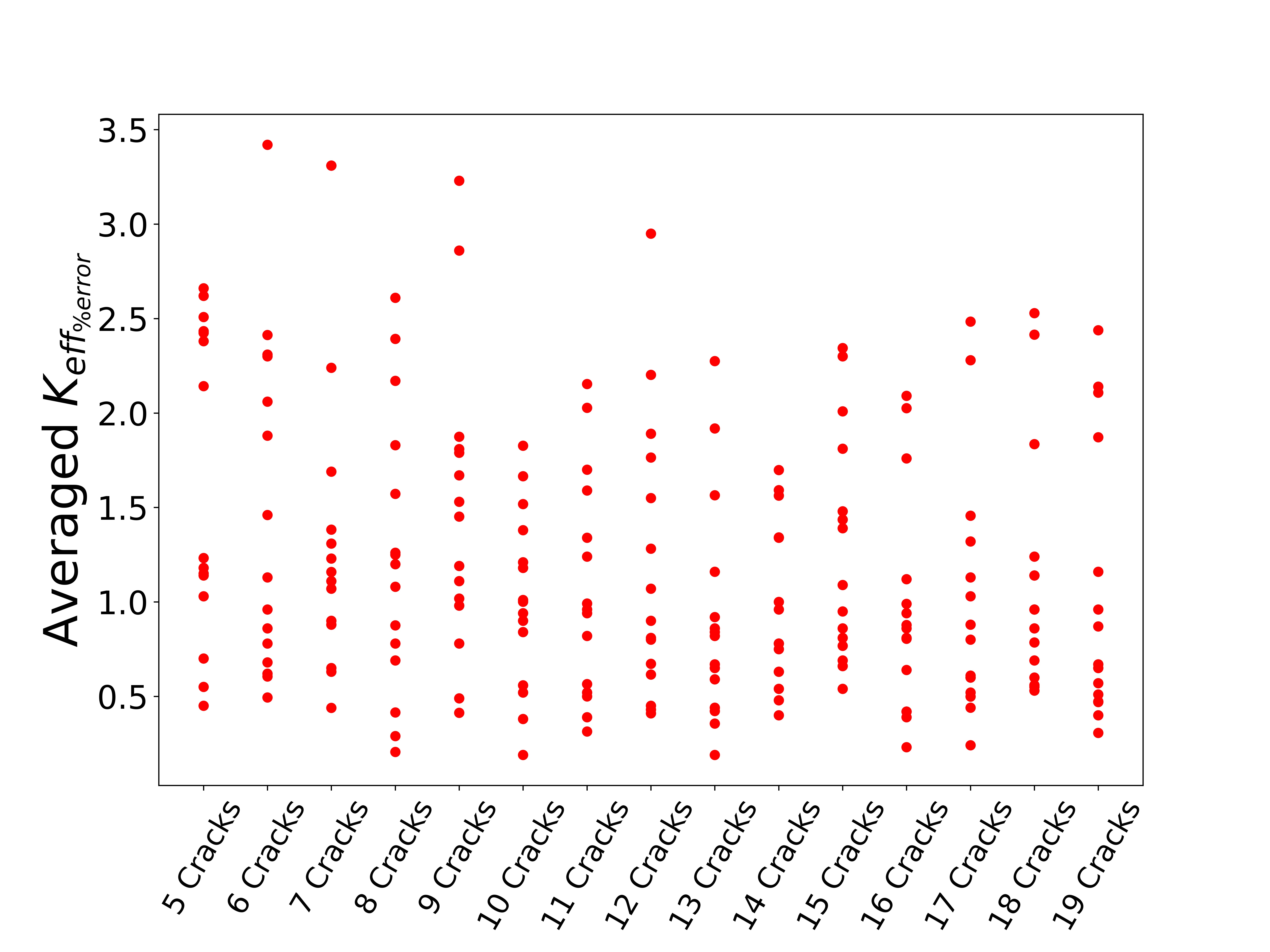}
            \centering
            \caption{Averaged maximum percentage errors of effective stress intensity factors errors for all test cases (225 test cases).}
            \label{fig:Keff_testset_error}
        \end{figure}

        Following the approach described in Section \ref{subsec:results_Keff},we computed the maximum percent errors in the predicted effective stress intensity factors at each time-step for each simulation in the test set. 
        Next, for each simulation, we computed the average over time resulting in 10 error points for each $C$, as shown in Figure \ref{fig:Keff_testset_error}.  
        The resulting highest effective stress intensity factor error from all simulations in the test dataset can be noted for the case of 6 microcracks at approximately $3.48\%$ error.
    
        \begin{figure}
            \centering 
            \includegraphics[width=0.75\linewidth]{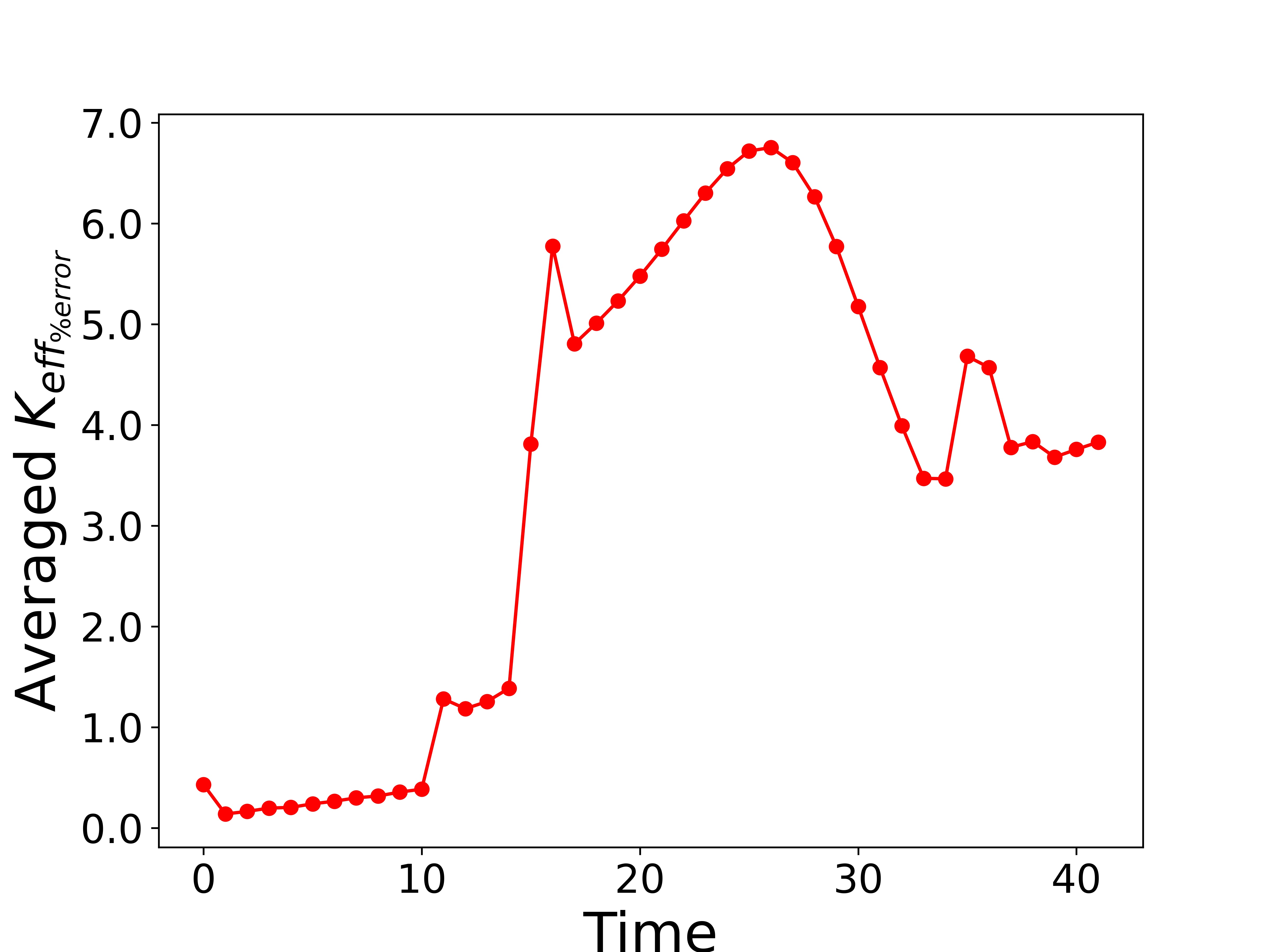}
            \centering
            \caption{Effective stress intensity factor maximum percentage error versus time for test 1 of 6 microcracks.}
            \label{fig:Keff_test1_6Cracks}
        \end{figure}
        
\begin{figure}
            \centering 
            \includegraphics[width=\linewidth]{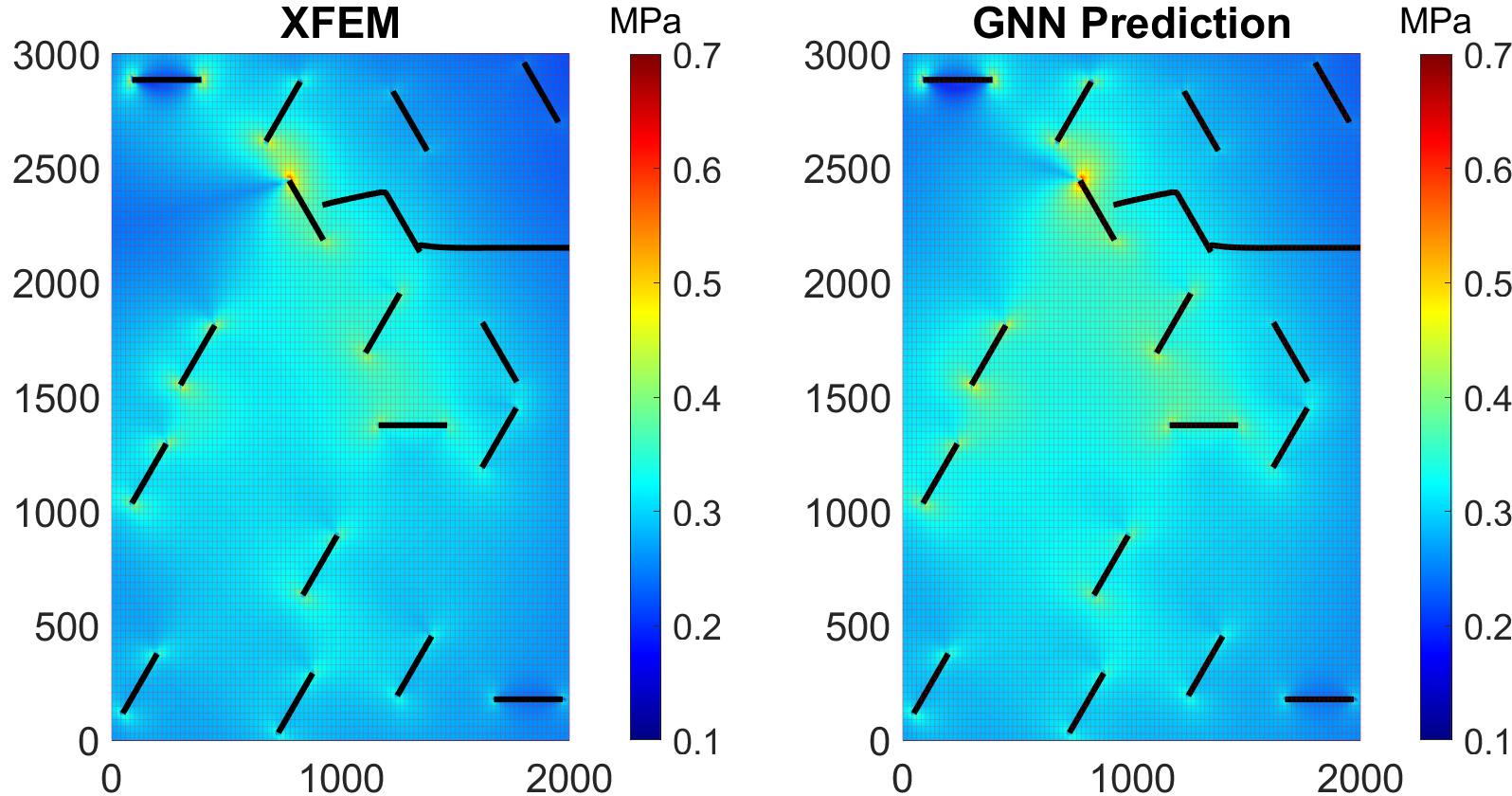}
            \centering
            \caption{Comparison of Von Mises stress distribution using XFEM (left) versus Microcrack-GNN (right) for test 1 of 6 microcracks time-step 26}
            \label{fig:SVM_Keff_comparison}
        \end{figure}

        To recognize the origin of this stress intensity factor error for the case of 6 microcracks, we look at Figures \ref{fig:Keff_test1_6Cracks} and \ref{fig:SVM_Keff_comparison}.
        Figure \ref{fig:Keff_test1_6Cracks} shows the time evolution of the error in stress intensity factor for test 1 of 6 microcracks.
        We note that during the initial stage of the simulation the errors remain below 2$\%$. 
        At approximately time-step 15, an increasing trend in the error is seen until reaching its peak of approximately 6.75$\%$ error during time-step 26.
        From Figure \ref{fig:SVM_Keff_comparison}, we present a qualitative comparison between Von Mises stress distribution (computed as described in Sections \ref{subsect:KI-GNN} and \ref{subsect:KI-GNN}) at time-step 26 for test 1 of $C=6$, as predicted by XFEM (left) and Microcrack-GNN (right).
        We note that while the error in the effective stress intensity factor reaches its maximum of 6.75$\%$ during this time-step, the Microcrack-GNN shows a very similar stress distribution to the XFEM model.

    \subsection{Von Mises Stress distribution error}
        
        One of the advantages of the Microcrack-GNN framework is its ability to predict stress distribution and evolution with time.
        As described in Sections \ref{subsect:KI-GNN} and  \ref{subsect:KII-GNN}, the predicted Mode-I and Mode-II stress intensity factors can be directly used to compute the stress distribution in the domain.
        
\begin{figure}
            \centering 
            \includegraphics[width=0.75\linewidth]{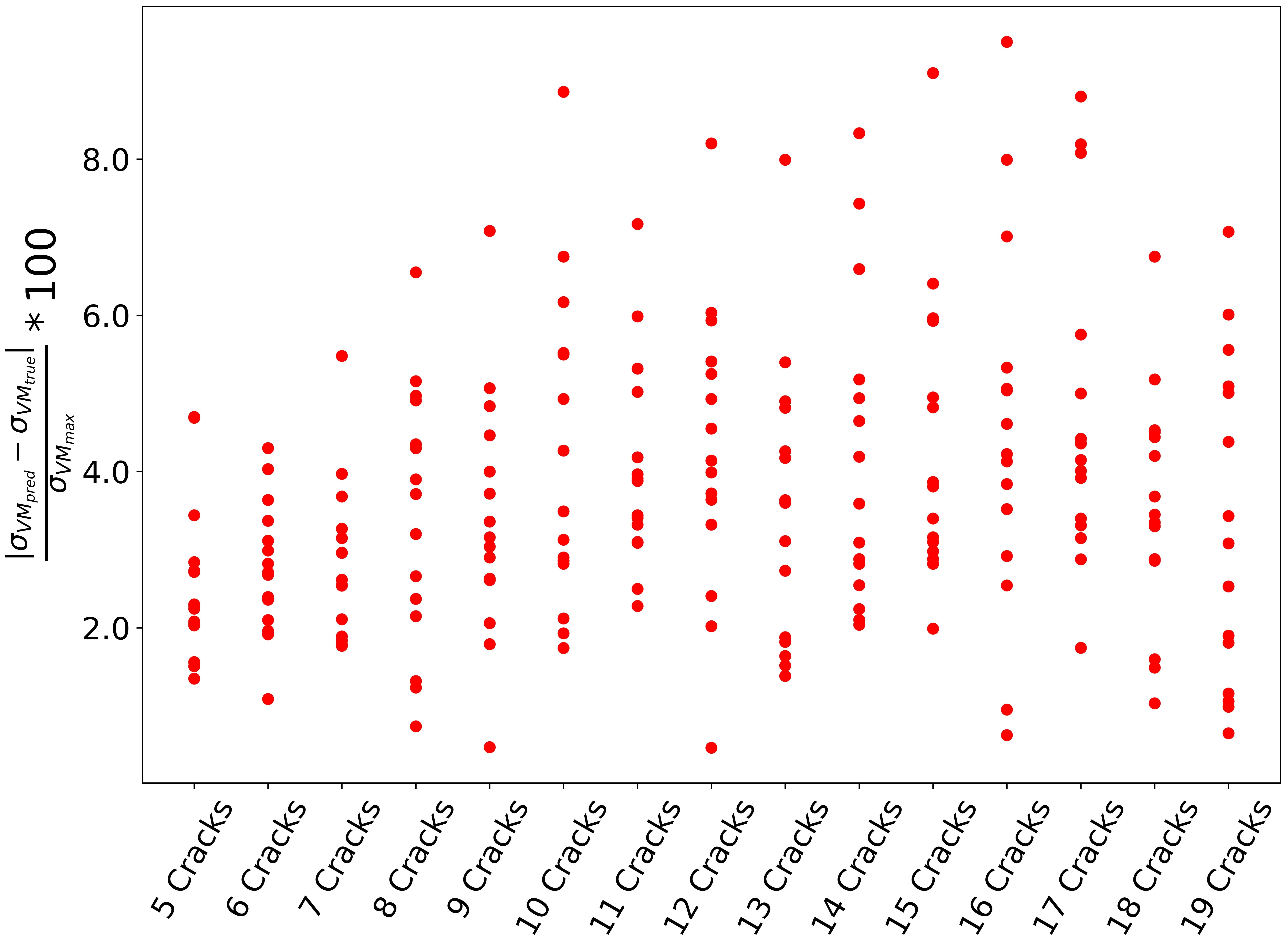}
            \centering
            \caption{Maximum percentage errors of Von Mises stresses for all test cases (225 test cases).}
            \label{fig:Sigma_testset_error}
        \end{figure}

        To show the resulting errors from the Von Mises stress prediction of the entire test dataset, we implemented the following approach.
        First, at each time-step we computed the absolute error (in MPa) for each point in the domain as $\| \sigma_{VM_{Pred}} - \sigma_{VM_{True}} \|$. 
        Then, we used the maximum Von Mises stress value ($\sigma_{{VM}_{Max}}$) at the corresponding time-step as the reference value for the error percentage as shown in equation (\ref{eq:Sigma_error}).
        \begin{flalign}
            {\sigma_{VM}}_{\% error} = \max_{(i,j)\in \mathbb{R}^{2}} \left( \frac{\left| {\sigma_{VM}}_{Pred} - {\sigma_{VM}}_{True} \right|}{{\sigma_{VM}}_{Max}} \right) \times 100
        \label{eq:Sigma_error}
        \end{flalign}
        Using this approach, we obtain the maximum percent error of Von Mises stress at each time-step for each simulation in the test dataset.
        Lastly, we compute the maximum Von Mises stress percent error across time for each test simulation as shown in Figure \ref{fig:Sigma_testset_error}.
        
        \begin{figure}
            \centering 
            \includegraphics[width=0.75\linewidth]{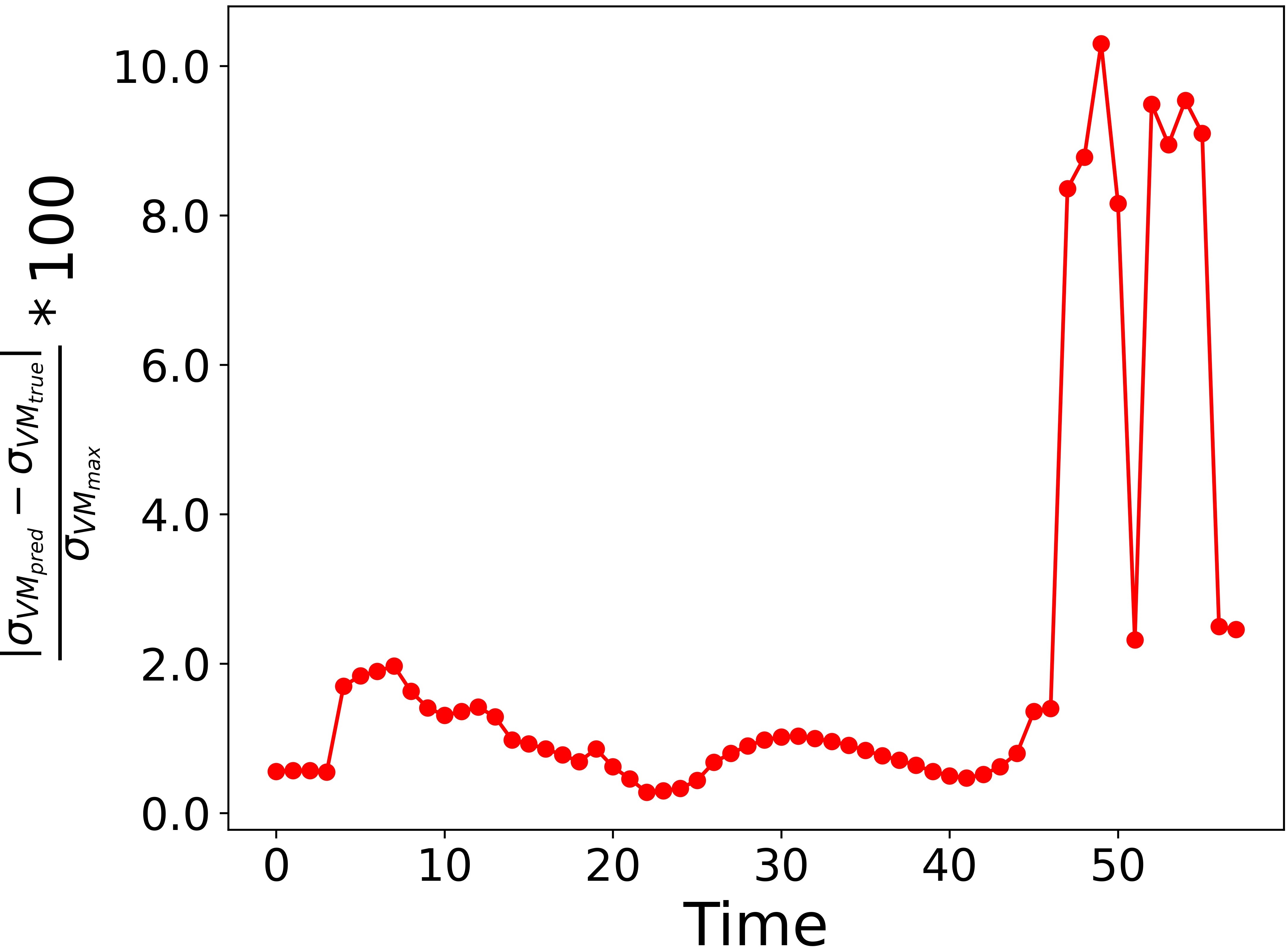}
            \centering
            \caption{Von Mises stress maximum percentage error versus time for test 6 of 16 microcracks.}
            \label{fig:Sigma_test6_16Cracks}
        \end{figure}
        
        \begin{figure}
            \centering 
            \includegraphics[width=\linewidth]{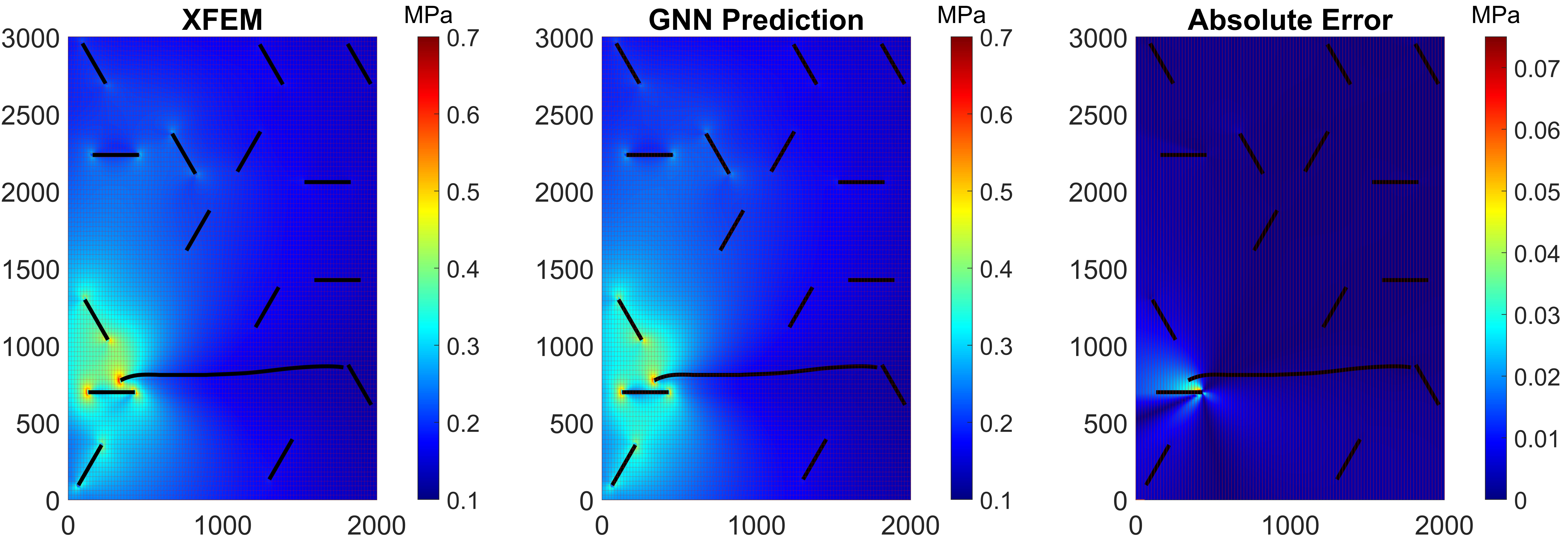}
            \centering
            \caption{Comparison of Von Mises stress distribution using XFEM (left-most) versus Microcrack-GNN (center), and resulting absolute error (MPa) (right-most) for test 6 of 16 microcracks}
            \label{fig:Dist_test6_16Cracks}
        \end{figure}
        
        Figure \ref{fig:Sigma_testset_error} shows that the Microcrack-GNN framework generated errors above $8\%$ in cases involving 10, 12, 14, 15, 16, and 17 microcracks.
        To understand the origin of these high errors, we look at Figure \ref{fig:Sigma_test6_16Cracks} showing the maximum percentage error of Von Mises stress at each time-step for test case 6 of 16 microcracks.
        We chose this test case as it showed the highest percent error of Von Mises stress {($10.29\%$)} from the entire dataset as shown in Figure \ref{fig:Sigma_testset_error}.
        From Figure \ref{fig:Sigma_test6_16Cracks}, we see that the maximum percent error stays below $3\%$ error throughout the majority of the simulation. However, at approximately time-step 47 the error increases rapidly above $8\%$ error and reaches its peak of {$10.29\%$} error at time-step 49.
        Figure \ref{fig:Dist_test6_16Cracks} shows the Von Mises stress distribution for this test case at time-step 49 (where maximum error occurs) resulting from the XFEM-based model (left-most), the Microcrack-GNN framework (center), and the absolute error between the two (right).
        The source of error can be seen to originate from the stress intensity factors of the cracks coalescing.
        Therefore, one of the limitations of the developed framework is the high error resulting when two or more cracks coalescence.
        Since the Von Mises stresses are computed using the predicted Mode-I and Mode-II stress intensity factors, which are then superimposed for all crack-tips in the system, the error adds up.
        A possible future work to circumvent this challenge is to predict the Von Mises stress directly throughout the mesh.

\end{document}